\documentclass[twocolumn,aps,prc,amsmath,showpacs,floatfix,nofootinbib]
{revtex4-1}

\usepackage{CJK}                     
\usepackage[dvips]{graphicx}
\usepackage{bm}                      
\usepackage{mathptmx}                
\usepackage{dcolumn}                 
\usepackage{xcolor}

\def\bea{\begin{eqnarray}}
\def\eea{\end{eqnarray}}
\def\be{\begin{equation}}
\def\ee{\end{equation}}
\def\rra{\right\rangle}
\def\lla{\left\langle}

\def\rv{\bm{r}}
\def\kv{\bm{k}}
\def\eps{\varepsilon}
\def\esym{S}

\begin{document}

\title{
Nuclear matter equation of state from a quark-model nucleon-nucleon interaction}

\begin{CJK*}{GB}{gbsn} 

\author{
K. Fukukawa,$^{1,2}$ M. Baldo,$^1$ G. F. Burgio,$^1$ L. Lo Monaco,$^3$
and H.-J. Schulze$^1$}
\affiliation{
$^1$ INFN Sezione di Catania, Dip.~di Fisica, Universit\`a di Catania,
Via Santa Sofia 64, I-95123 Catania, Italy}
\affiliation{
$^2$ Research Center for Nuclear Physics, Osaka University,
10-1 Mihogaoka, Osaka 567-0047, Japan}
\affiliation{
$^3$ Dipartimento di Fisica, Universit\`a di Catania,
Via Santa Sofia 64, I-95123 Catania, Italy}

\date{\today}

\begin{abstract}
Starting from a realistic constituent quark model for the
nucleon-nucleon interaction,
we derive the equation of state (EOS) of nuclear matter
within the Bethe-Brueckner-Goldstone approach up to three-hole-line level,
without need to introduce three-nucleon forces.
To estimate the uncertainty of the calculations both the gap and the continuous
choices for the single-particle potential are considered and compared.
The resultant EOS is compatible with the phenomenological analysis on
the saturation point,
the incompressibility,
the symmetry energy at low density and its slope at saturation,
together with the high-density pressure extracted from flow data on
heavy ion collisions.
Although the symmetry energy is appreciably larger in the gap choice
in the high-density region,
the maximum neutron star masses
derived from the continuous-choice EOS and the gap-choice EOS
are similar and close to two solar masses,
which is again compatible with recent observational data.
Comparison with other microscopic EOS is presented and discussed.
\end{abstract}


\maketitle
\end{CJK*}

\section{Introduction}

The nuclear matter equation of state (EOS) is one of the central issues
in nuclear physics.
Its detailed knowledge would allow us to connect the data obtained
in laboratory experiments on heavy ion collisions (HIC) and the processes
that characterize the structure and evolution of compact astrophysical objects
like neutron stars (NS) and supernovae.
On the other hand, laboratory experiments and astrophysical observations
can put meaningful constraints on the nuclear EOS.
Unfortunately a direct link between phenomenology and the EOS is not possible
and theoretical frameworks and inputs are necessary for the interpretation
of the data.
In particular the EOS above saturation density is much less constrained
than around or below saturation.

An intense activity, lasting since several years,
has been developed in analyzing and interpreting
the experimental and observational data
for the purpose of putting severe constraints on the EOS
and on the corresponding theoretical models \cite{RPP}.
It can be recognized from these efforts
that a sound theoretical and microscopical framework for modeling the EOS
can be of help for the establishment of firm results on the EOS properties.
Along these lines a microscopic many-body theory based on interactions
among nucleons,
that stems from strong interaction theory,
can be of great relevance in reducing the uncertainties that characterize this
type of analysis.

Meson-exchange models of the nucleon interaction
have been extensively developed since several years,
and applied to nuclear matter and NS structure within many-body theory.
One can mention the variational method \cite{APR},
the relativistic Dirac-Brueckner-Hartree-Fock (DBHF) \cite{DBHF},
and the non-relativistic Bethe-Brueckner-Goldstone (BBG) expansion
\cite{Day,Song,Bal1,sartor},
which have employed different versions of nucleon-nucleon interactions
inspired by the meson-exchange model.
In the non-relativistic scheme three-body forces (TBF) have been introduced
to obtain the correct saturation point of nuclear matter \cite{Urb,Tar,Umb,Hans}.
The ambition of all these approaches is to devise an elementary interaction
among nucleons that is able to describe both few-body nuclear systems
and nuclear matter in agreement with the existing phenomenological data.
This program has been only partially successful.
It turns out in fact that it is difficult to reproduce the binding
energy of three- and four-body systems and at the same time
to predict the correct saturation point within this scheme.

More recently the chiral expansion theory to the nucleon interaction
has been extensively developed
\cite{Weinberg,Ent,Valder,Leut,Ulf,Epel,Ca,Hebel,Diri,Hebel2,Rios,Mac,MacCor}.
This approach is based on a deeper level of the strong interaction theory,
where QCD chiral symmetry is explicitly exploited in a low-momentum expansion
of the multi-nucleon interaction processes.
In this approach multi-nucleon interactions arise naturally and a hierarchy of
the different orders can be established.
Despite some ambiguity in the parametrization of the force \cite{Mac}
and some difficulty in the treatment of many-body systems \cite{Rupak},
the method has marked a great progress in the microscopic theory
of nuclear systems.
Indeed it turns out \cite{MacCor,Ekst} that within this class of interactions
a compatible treatment of few-nucleon systems and nuclear matter is possible.
Along the same lines a chiral force \cite{Lynn} has been adjusted to reproduce,
within a Monte Carlo calculation,
the binding of $^4$He and the phase shifts of neutron-alpha scattering.
The same interaction was used to describe neutron matter.
Coupled cluster calculations with chiral forces including three-body forces
have been performed for neutron and symmetric nuclear matter \cite{Hagen,Baar},
and finite nuclei \cite{Bind}.

Another approach inspired by the QCD theory of strong interaction has been
developed by a few groups \cite{OY8081,Wong86,Oka87,Shm89,Shm00,Var05,QMPPNP}.
In this approach the quark degree of freedom is explicitly introduced
and the nucleon-nucleon interaction is constructed from gluon and meson exchange
between quarks,
the latter being confined inside the nucleons.
One of these quark models of the nucleon-nucleon interaction,
named fss2 \cite{QMPPNP,fss2NN},
is able to reproduce closely the experimental phase shifts
and the few-body binding energies \cite{fss2NN,hypt08,ndscatreview,QMalpha}.
More recently it has been shown \cite{PRL2014} that the fss2 interaction
is able to reproduce also the correct nuclear matter
saturation point without any additional parameter
or need to introduce TBF.

In this paper we analyze further the fss2 interaction.
On one hand we compare the results in nuclear matter
with additional phenomenological constraints,
on the other hand we extend the EOS based on the fss2 interaction
to higher density and apply it to NS calculations.
In this study we use the renormalized energy-independent kernel
of this model \cite{ndscat1}.
Kernels of quark-model nucleon-nucleon interactions are obtained
by using the resonating-group method (RGM) for the $(3q)$-$(3q)$ system.
Although they are therefore energy dependent,
we can eliminate the energy dependence,
as we will see in the next section.

This paper is organized as follows.
In Section~\ref{s:fss} we introduce the quark-model nucleon-nucleon interaction
fss2 \cite{fss2NN,fss2YN} and the formulation for the renormalized RGM kernel.
We will briefly review phase shifts and deuteron properties.
The explicit form of the deuteron wave function will be given
in Appendix~\ref{s:deut}.
In Section~\ref{s:bhf} we first recapitulate the Brueckner-Hartree-Fock (BHF)
calculation and the Bethe-Faddeev calculation,
based on the BBG framework \cite{book}.
Then the nuclear matter EOS for the fss2 interaction is reported and its
properties are discussed in relation to the phenomenological constraints,
also in comparison with some other theoretical methods and interactions.
Section~\ref{s:ns} is devoted to the calculations of NS structure.
Conclusions are drawn in Section~\ref{s:end}.

\section{Quark-model baryon-baryon interaction fss2}
\label{s:fss}

The fss2 baryon-baryon interaction \cite{fss2NN,fss2YN}
is a low-energy effective model,
which introduces some essential features of QCD.
The color degree of freedom is explicitly considered
within the spin-flavor $SU(6)$ approximation
and the antisymmetrization of quarks is exactly taken into account
within the framework of RGM.
The full model Hamiltonian for the $(3q)$-$(3q)$ system reads
\bea
 H &=& \sum^6_{i=1} \left( m_i + \frac{p_i^2}{2m_i} \right) - T_G
\nonumber\\
 &+& \sum^6_{i<j} \left( U^\text{Cf}_{ij} + U^\text{FB}_{ij}
 + U^\text{S}_{ij} + U^\text{PS}_{ij} + U^\text{V}_{ij} \right) \:,
\eea
where $m_i$ and $p_i$ are the constituent quark mass and momentum
of each particle respectively,
and $T_G$ denotes the center-of-mass motion.
The remaining terms denote the effective quark-quark interaction.

The confinement potential $U^\text{Cf}_{ij}$
is a phenomenological $r^2$-type potential,
which has the favorable feature that it does not contribute
to the baryon-baryon interactions.
We use a color analogue of the Fermi-Breit (FB) interaction
$U^\text{FB}_{ij}$ with explicit quark-mass dependence,
motivated by the dominant one-gluon exchange process
in conjunction with the asymptotic freedom of QCD.
This includes the short-range repulsion and spin-orbit force,
both of which are successfully described.
On the other hand,
the medium-range attraction and the long-range tensor force,
especially those mediated by pions,
are extremely nonperturbative from the viewpoint of QCD.
These are therefore most relevantly described by effective
meson-exchange potentials.
Compared with the former version FSS \cite{FSS,FSSscat},
in which the scalar (S) and the pseudo-scalar (PS) nonets are included,
the introduction of the vector (V) nonets
and the momentum-dependent Bryan-Scott term \cite{Bry67}
greatly improves nucleon-nucleon phase shifts \cite{fss2NN}
and makes fss2 sufficiently realistic.

The RGM equation for the relative wave function $\chi(\rv)$ is given by
\be
 \lla \phi(3q)\phi(3q)|E-H|{\cal A}\{\phi(3q)\phi(3q)\chi(\rv)\} \rra
  = 0 \:,
\label{RGM}
\ee
where $\phi(3q)$ is the three-quark cluster (nucleon) wave function
and is described by a $(0s)^3$ harmonic oscillator
with a common width parameter.
The antisymmetrization operator is denoted by ${\cal A}$.
In our actual calculation,
Eq.~(\ref{RGM}) is solved in momentum space \cite{LS-RGM}.
We rewrite Eq.~(\ref{RGM}) in Schr{\"o}dinger-like form as
\be
 \left[ \eps - H_0 - V_\text{RGM}(\eps) \right] \chi(\rv)=0 \:,
\label{ene-dep}
\ee
where $\eps$ is the two-nucleon energy measured from its threshold
in the center-of-mass system and $H_0$ is the kinetic energy operator.
We regard $V_\text{RGM}(\eps)=V_D+G+\eps K$
as the non-local and energy-dependent potential.
Here, $V_D$ is the direct meson-exchange kernel,
$G$ represents all exchange kernels for the kinetic-energy
and interaction terms,
and $K$ is the exchange normalization kernel.

In the many-body scattering problem,
an energy-independent potential is desirable,
since the energy of a two-nucleon pair is not well-defined
in the in-medium scattering state.
The energy dependence of the RGM kernel can be reduced by renormalizing
the RGM kernel in the following way \cite{Suz08}.
We can rewrite Eq.~(\ref{ene-dep}) as
\be
 \left[\eps-N^{-1/2}\left(H_0+V_D+G \right)N^{-1/2} \right] \Psi(\rv) = 0 \:,
\label{ene-indep}
\ee
where
\be
 N \equiv \lla \phi(3q)\phi(3q)|{\cal A}|\phi(3q)\phi(3q) \rra \:
\ee
is the normalization kernel and
$\Psi(\rv) \equiv N^{1/2}\chi(\rv)$
is the renormalized RGM wave function.
If we define the non-local kernel
\be
 W \equiv N^{-1/2}(H_0+V_D+G)N^{-1/2} - (H_0+V_D+G) \:
\ee
and the renormalized RGM potential
$V^\text{RGM} \equiv V_D+G+W$,
then Eq.~(\ref{ene-dep}) becomes
\be
 \left[\eps-H_0-V^\text{RGM}\right] \Psi(\rv)=0 \:.
\ee
The detailed procedure to calculate $W$ can be found
in Appendix~A of Ref.~\cite{ndscat1}.
The asymptotic behavior of $\Psi(\rv)$ is the same as of $\chi(\rv)$,
because the square root of the normalization kernel approaches unity
at large distances.
The phase shifts derived from $\Psi(\rv)$ are the same
as those from $ \chi(\rv)$.

\def\myw{43mm}\def\mys{-1mm}\def\myv{-1mm}
\begin{figure*}[tbp]
\vspace{-10mm}
\includegraphics[width=\myw]{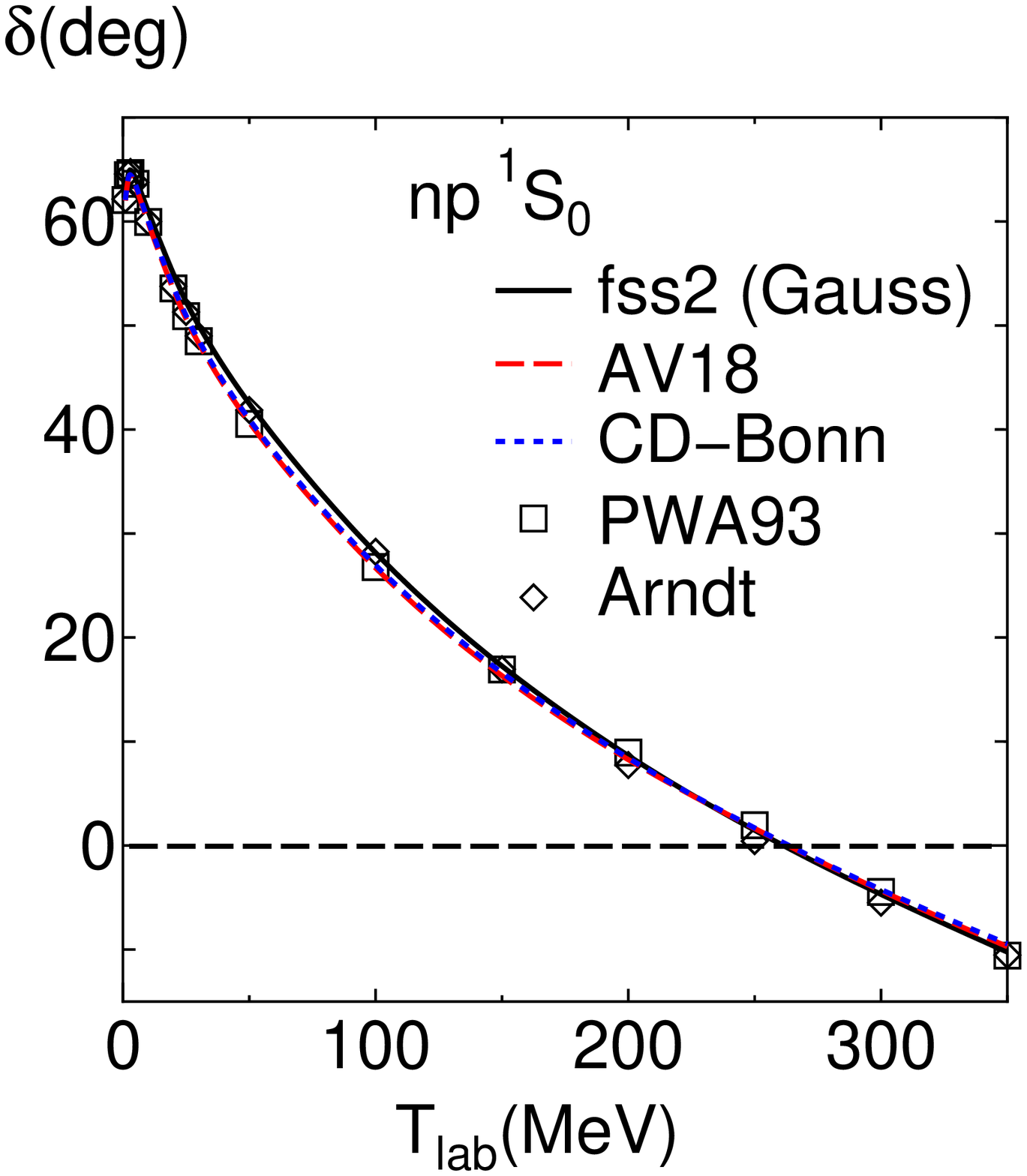}\hspace{\mys}
\includegraphics[width=\myw]{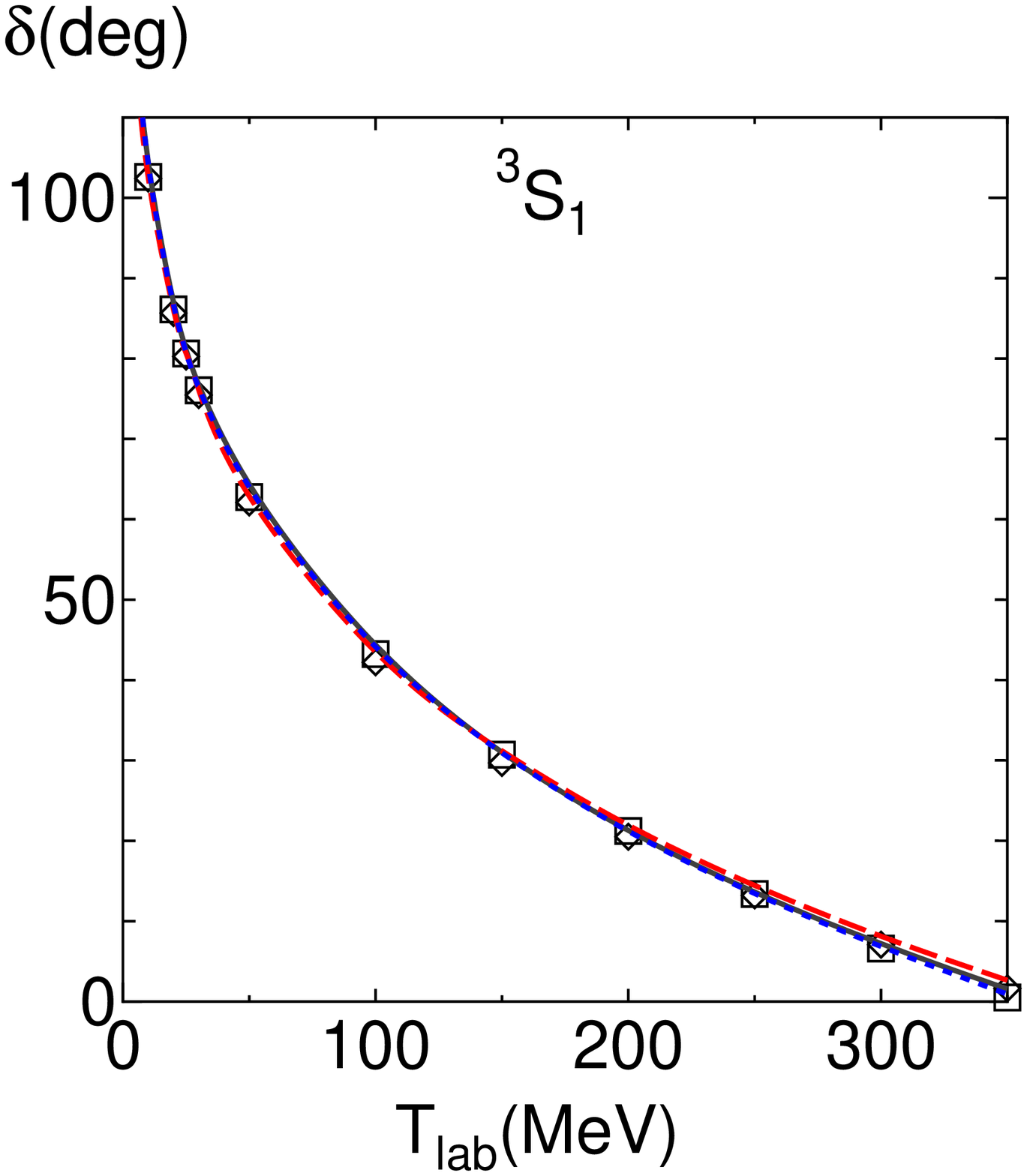}\hspace{\mys}
\includegraphics[width=\myw]{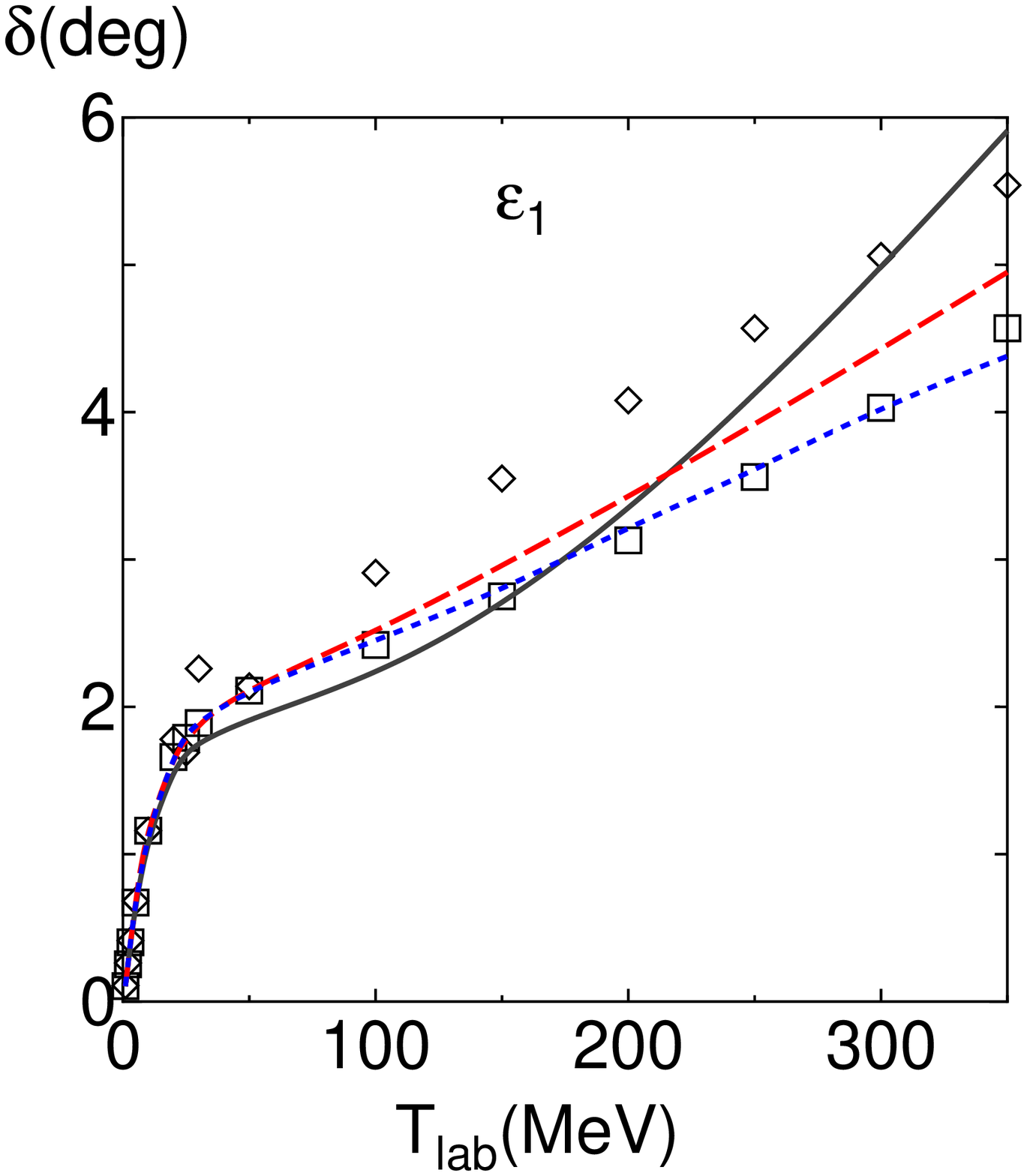} \hspace{\mys}
\includegraphics[width=\myw]{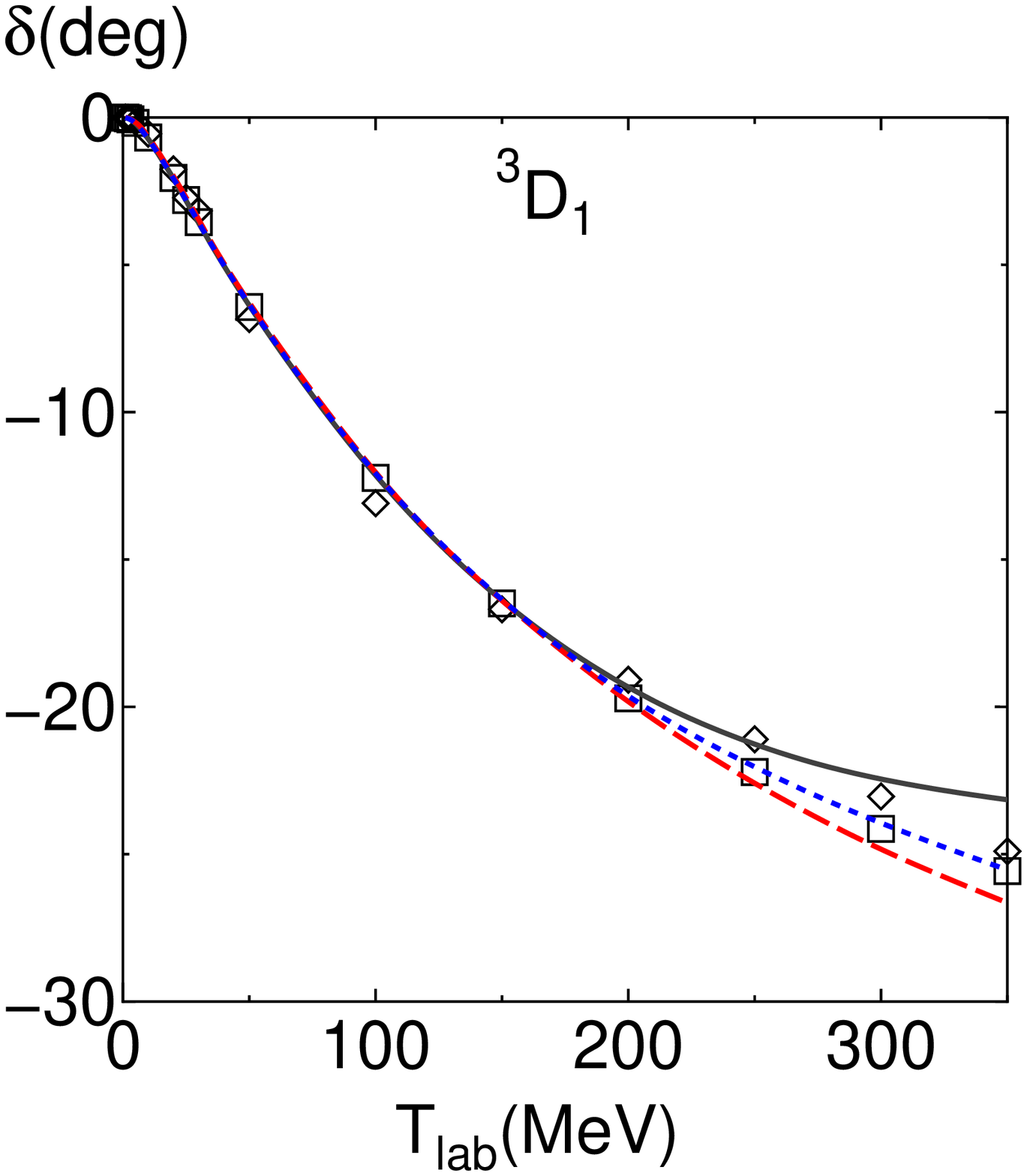}\vspace{\myv}

\includegraphics[width=\myw]{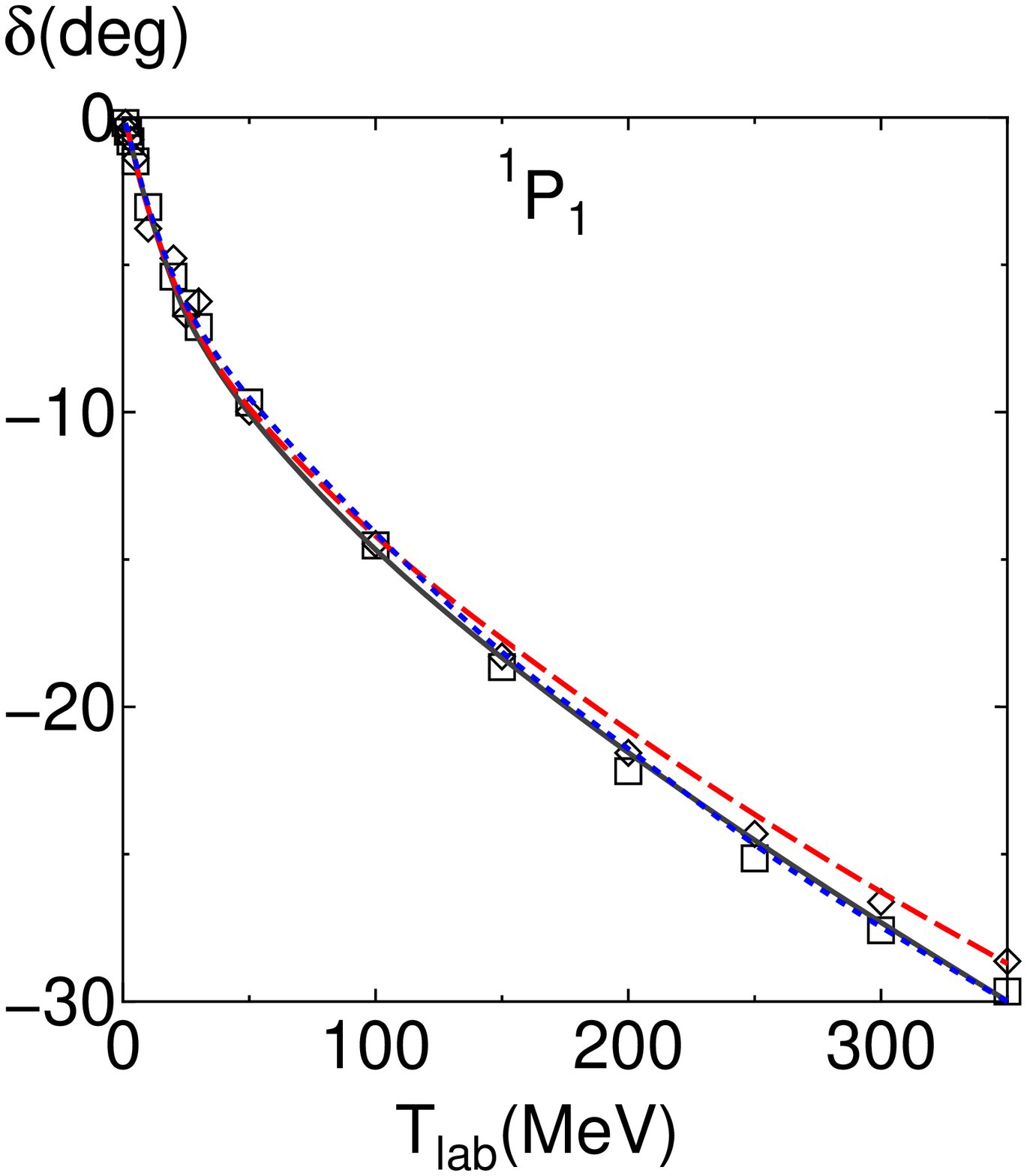}\hspace{\mys}
\includegraphics[width=\myw]{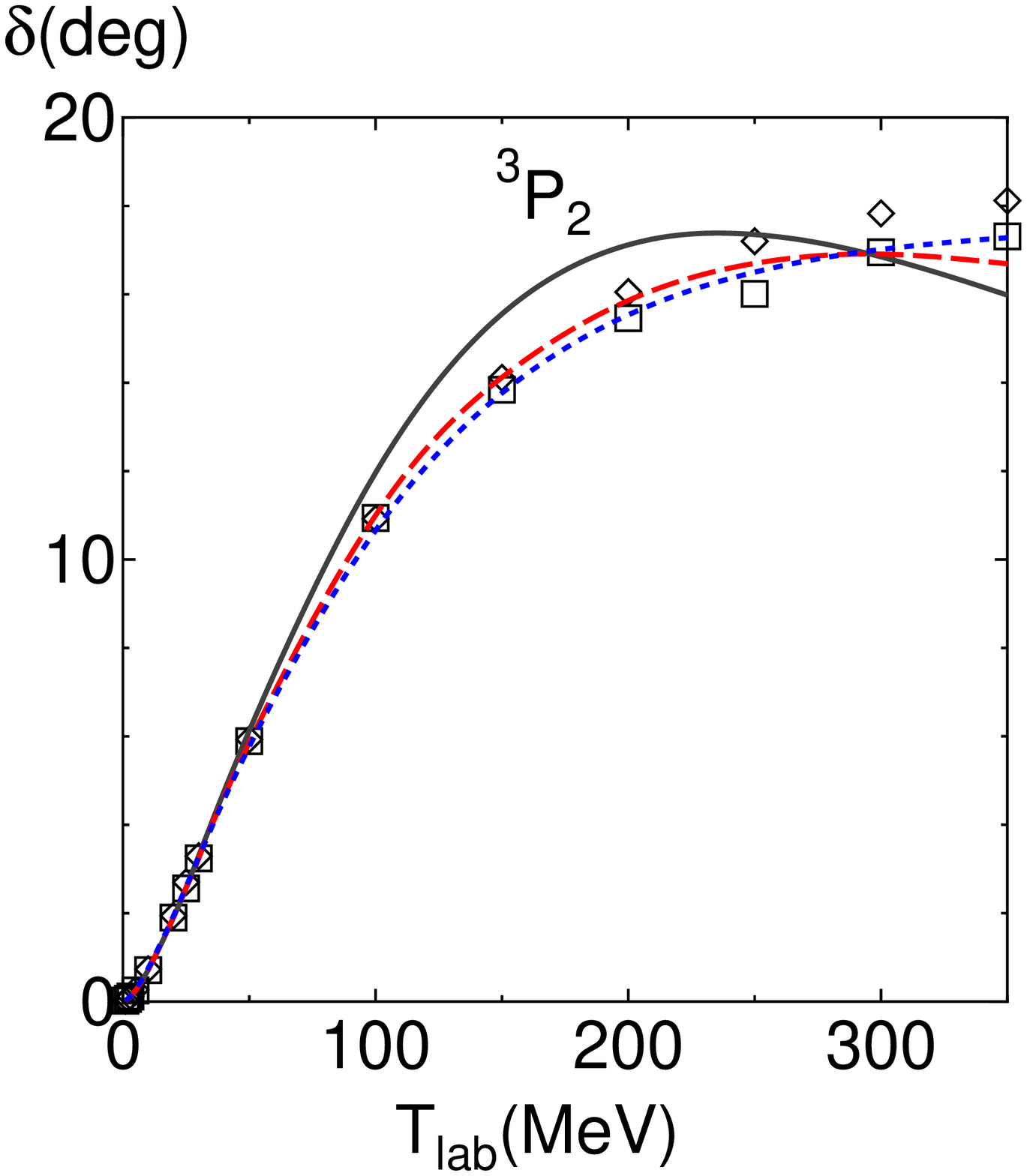}\hspace{\mys}
\includegraphics[width=\myw]{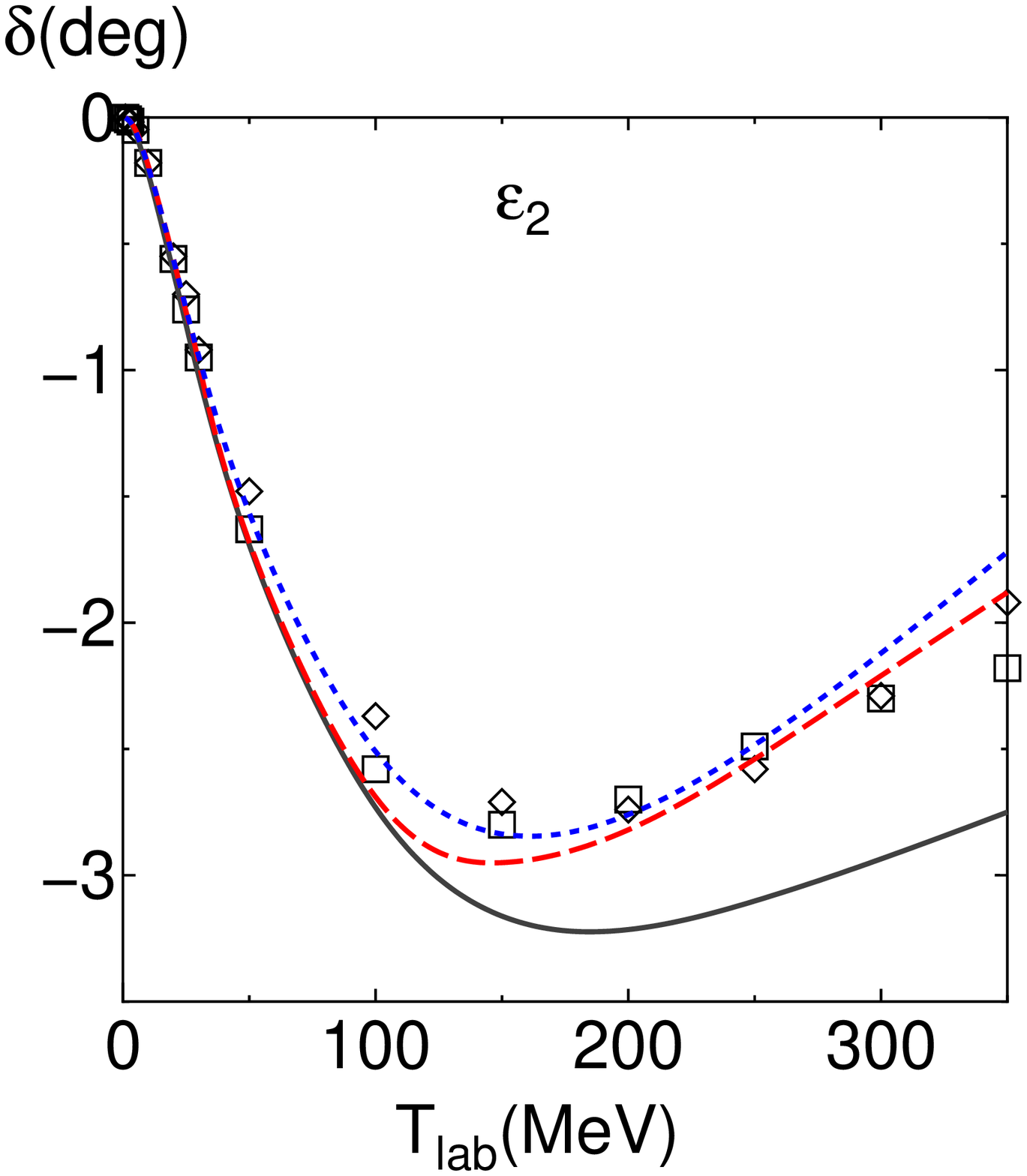} \hspace{\mys}
\includegraphics[width=\myw]{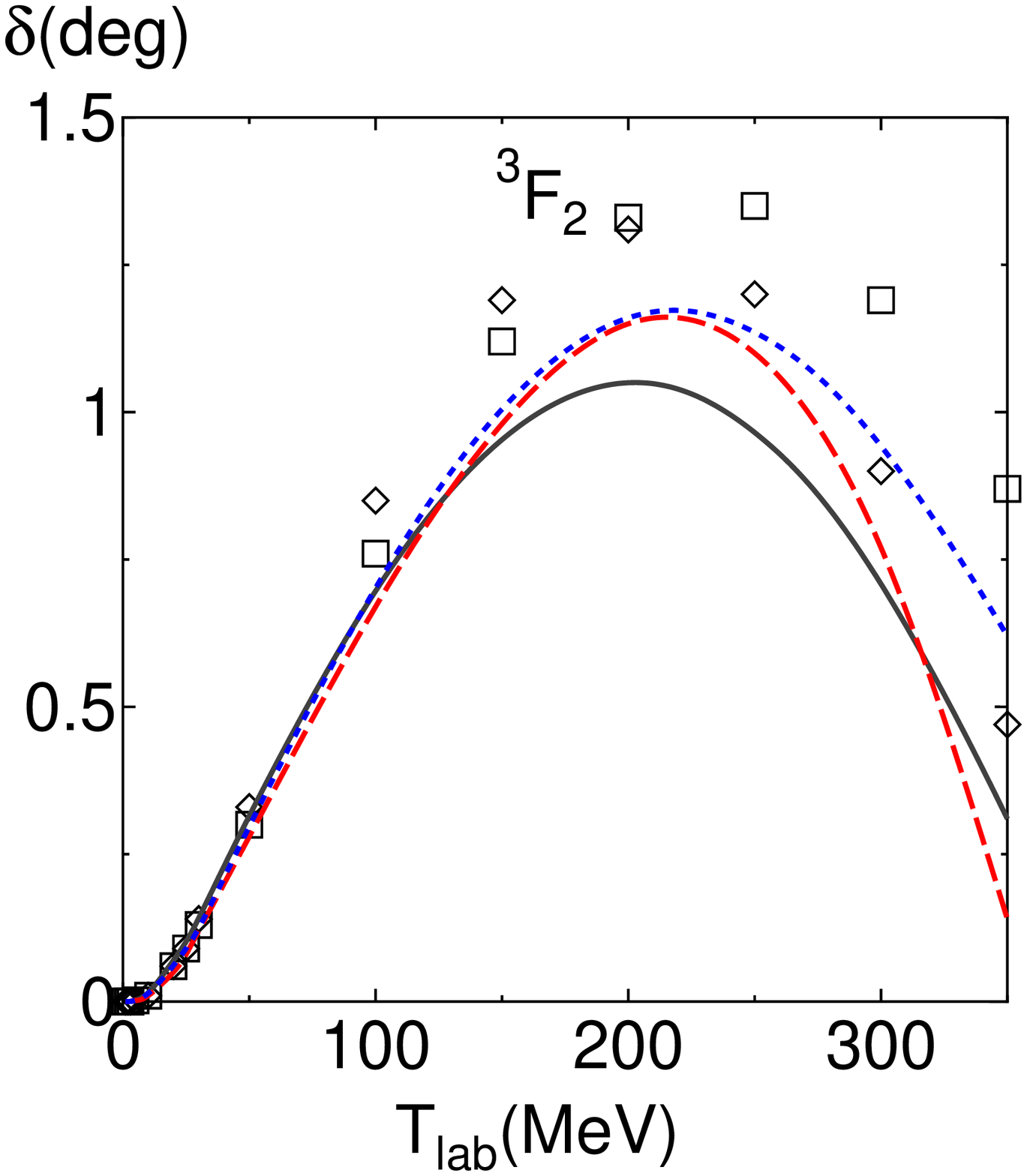}\vspace{\myv}

\includegraphics[width=\myw]{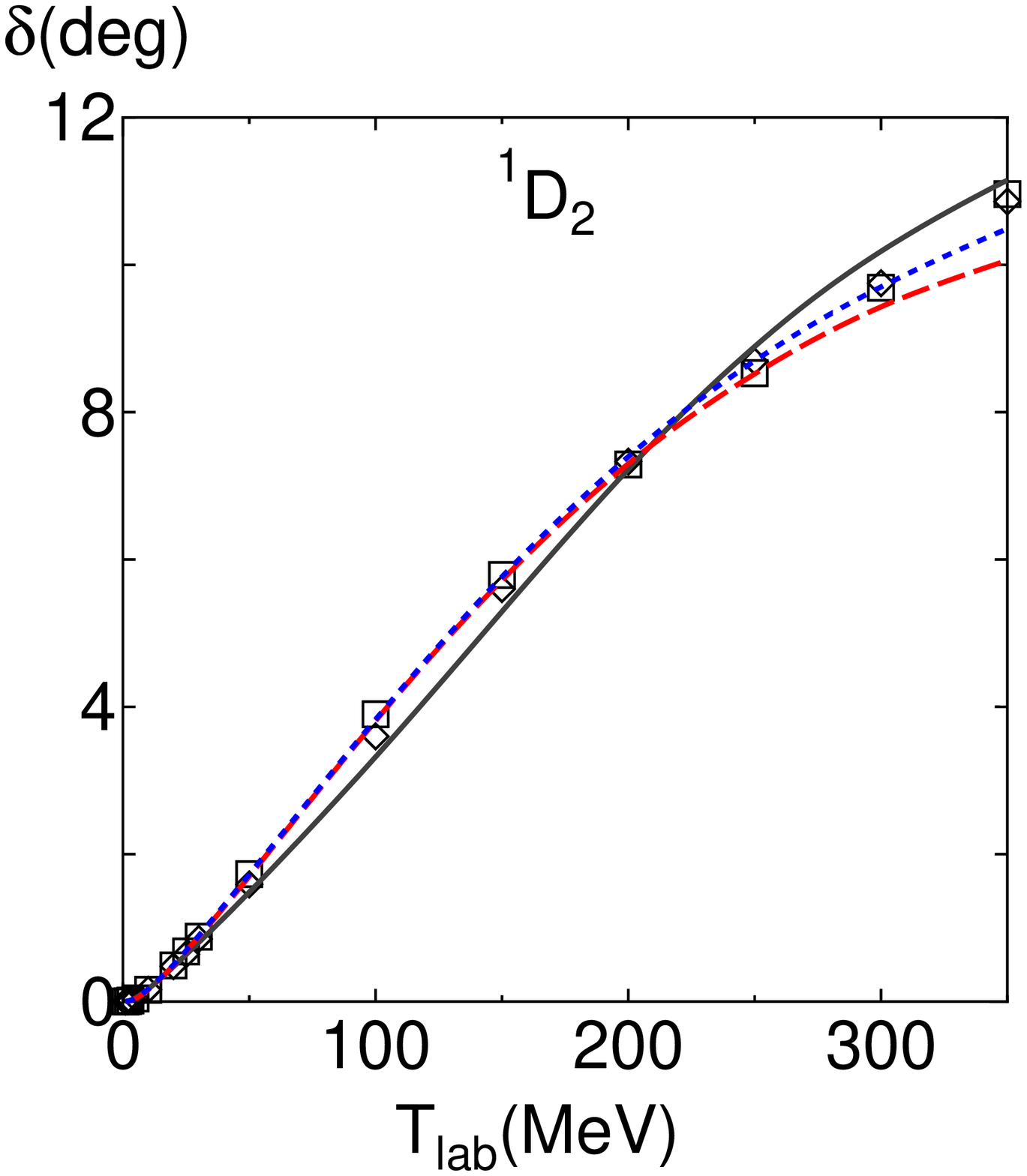}\hspace{\mys}
\includegraphics[width=\myw]{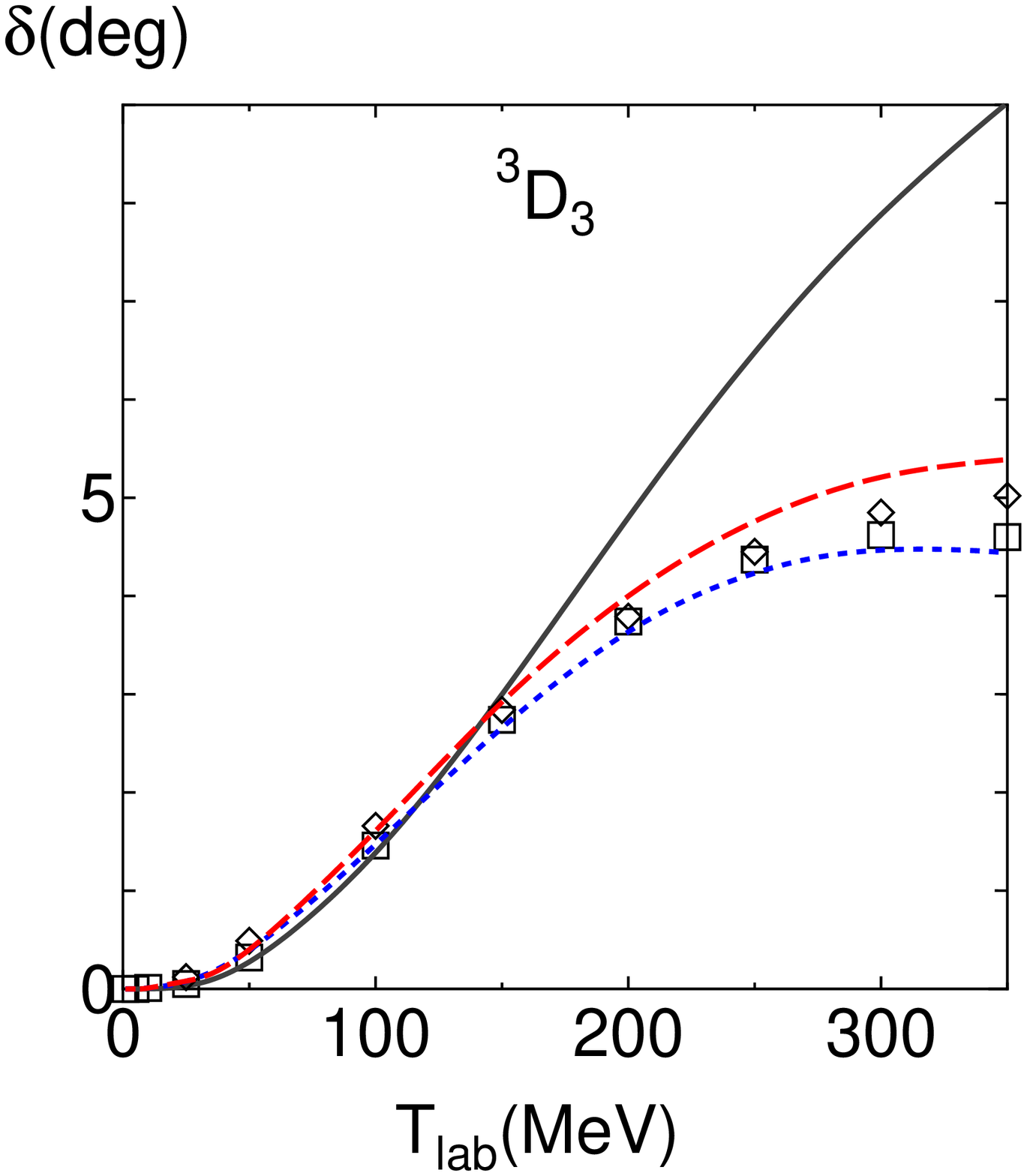}\hspace{\mys}
\includegraphics[width=\myw]{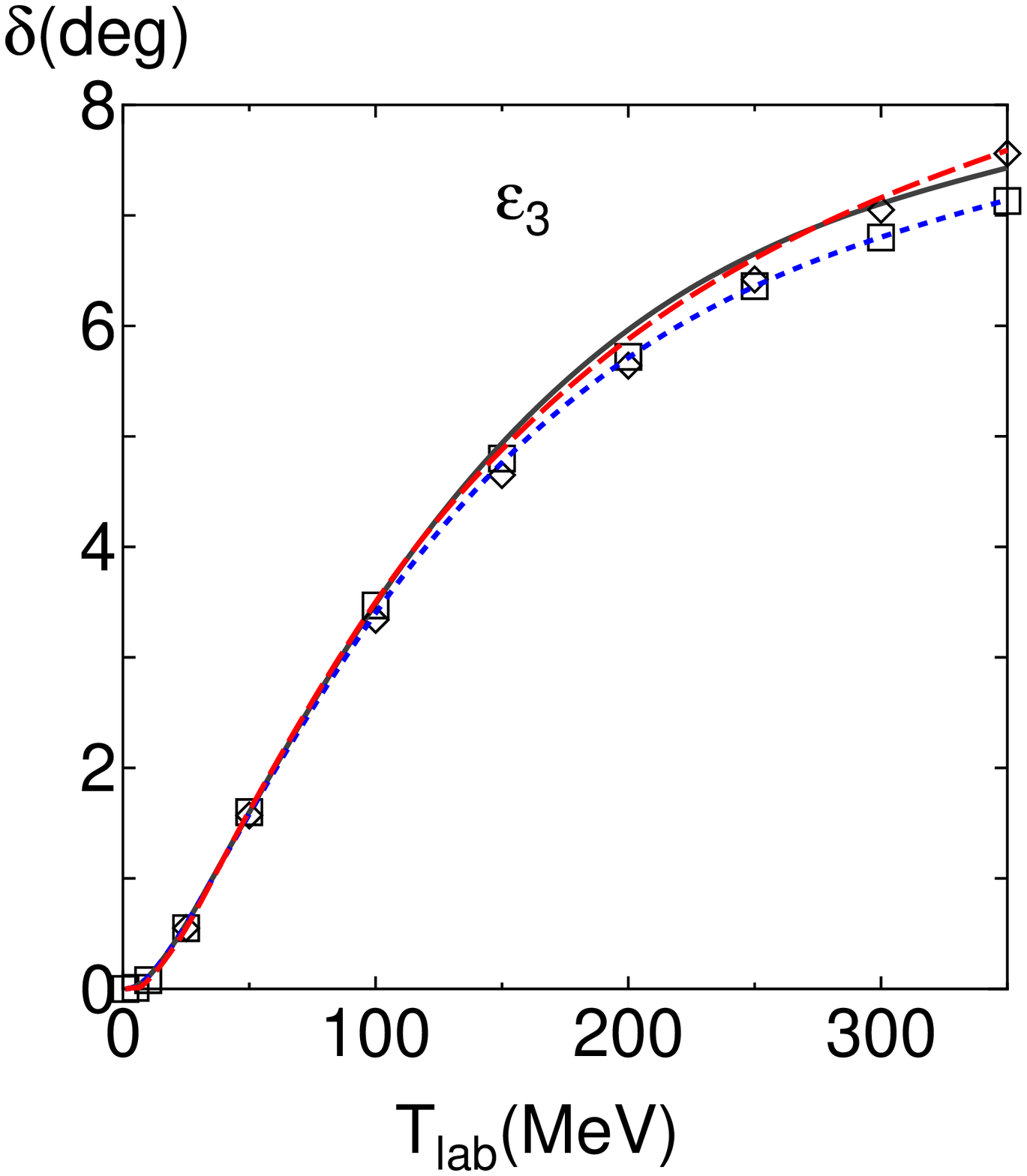} \hspace{\mys}
\includegraphics[width=\myw]{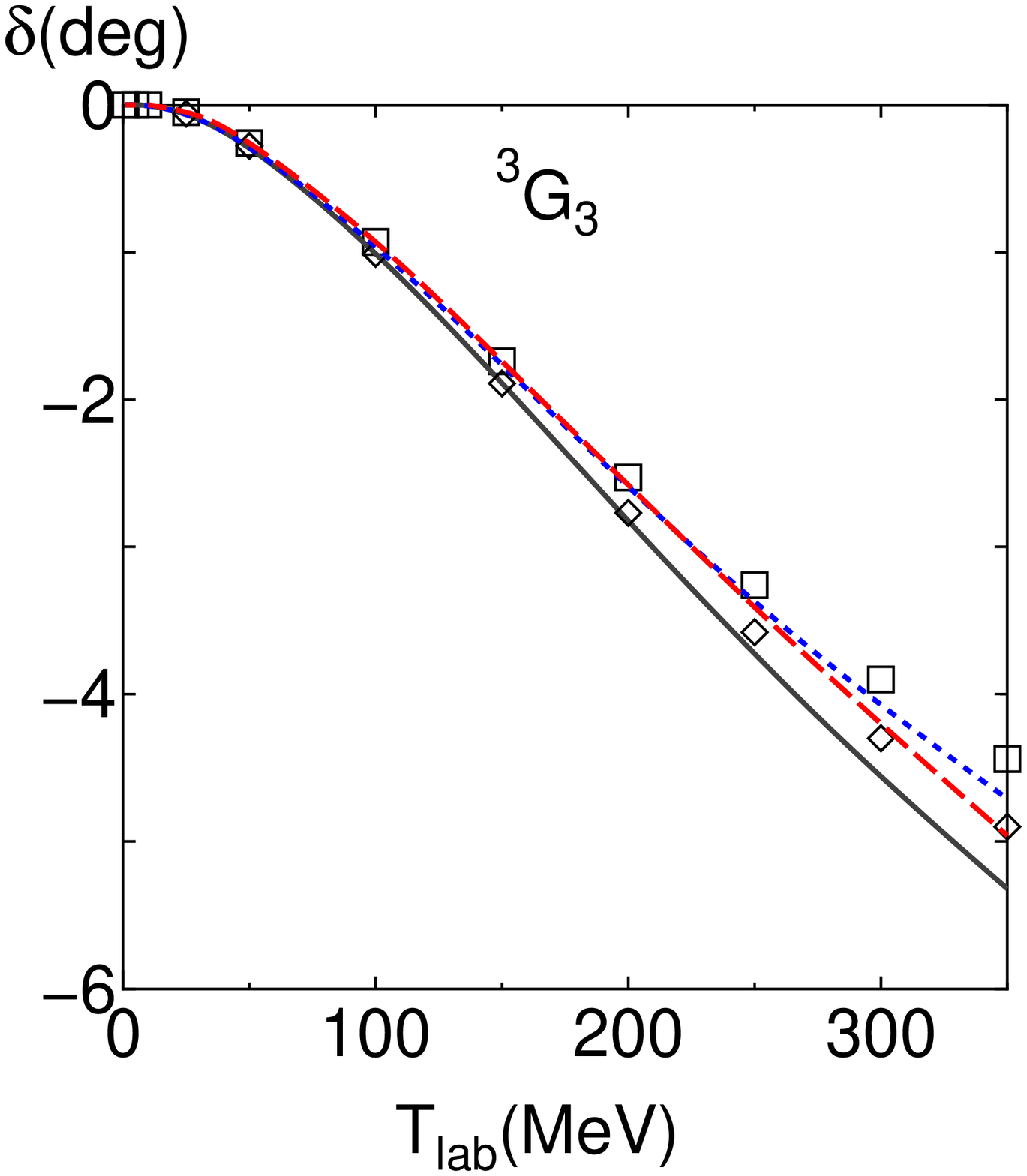}\vspace{\myv}

\includegraphics[width=\myw]{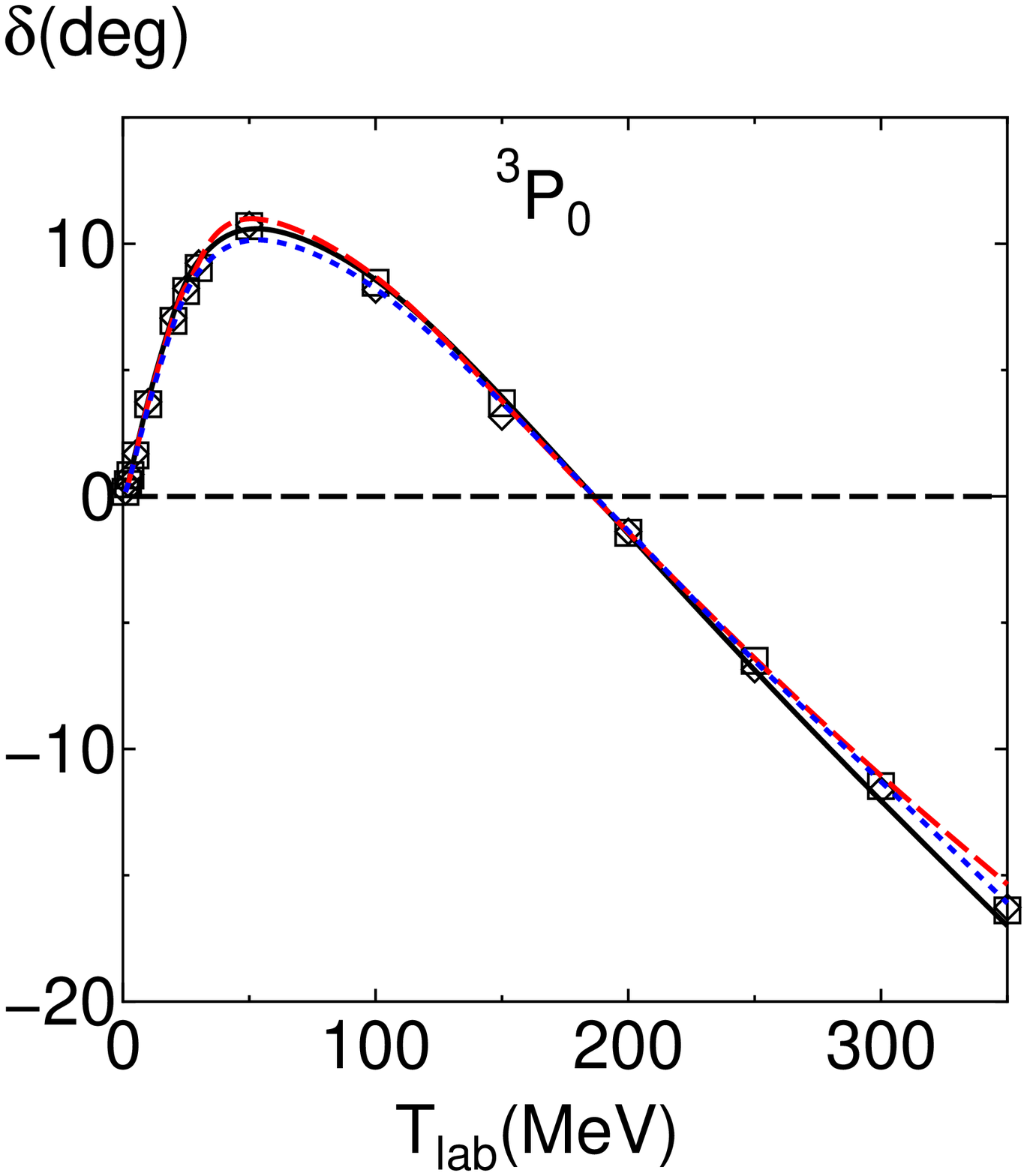}\hspace{\mys}
\includegraphics[width=\myw]{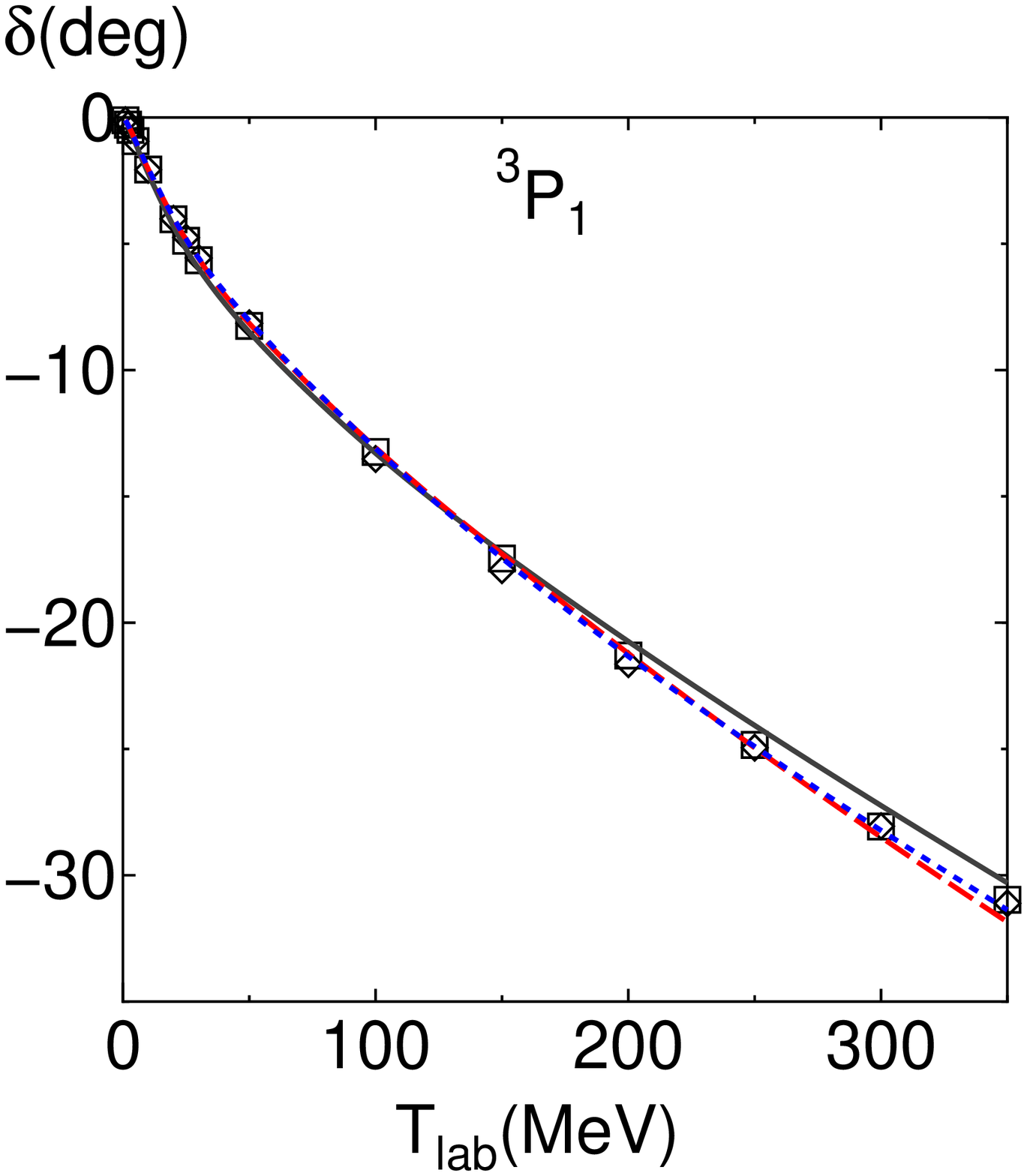}\hspace{\mys}
\includegraphics[width=\myw]{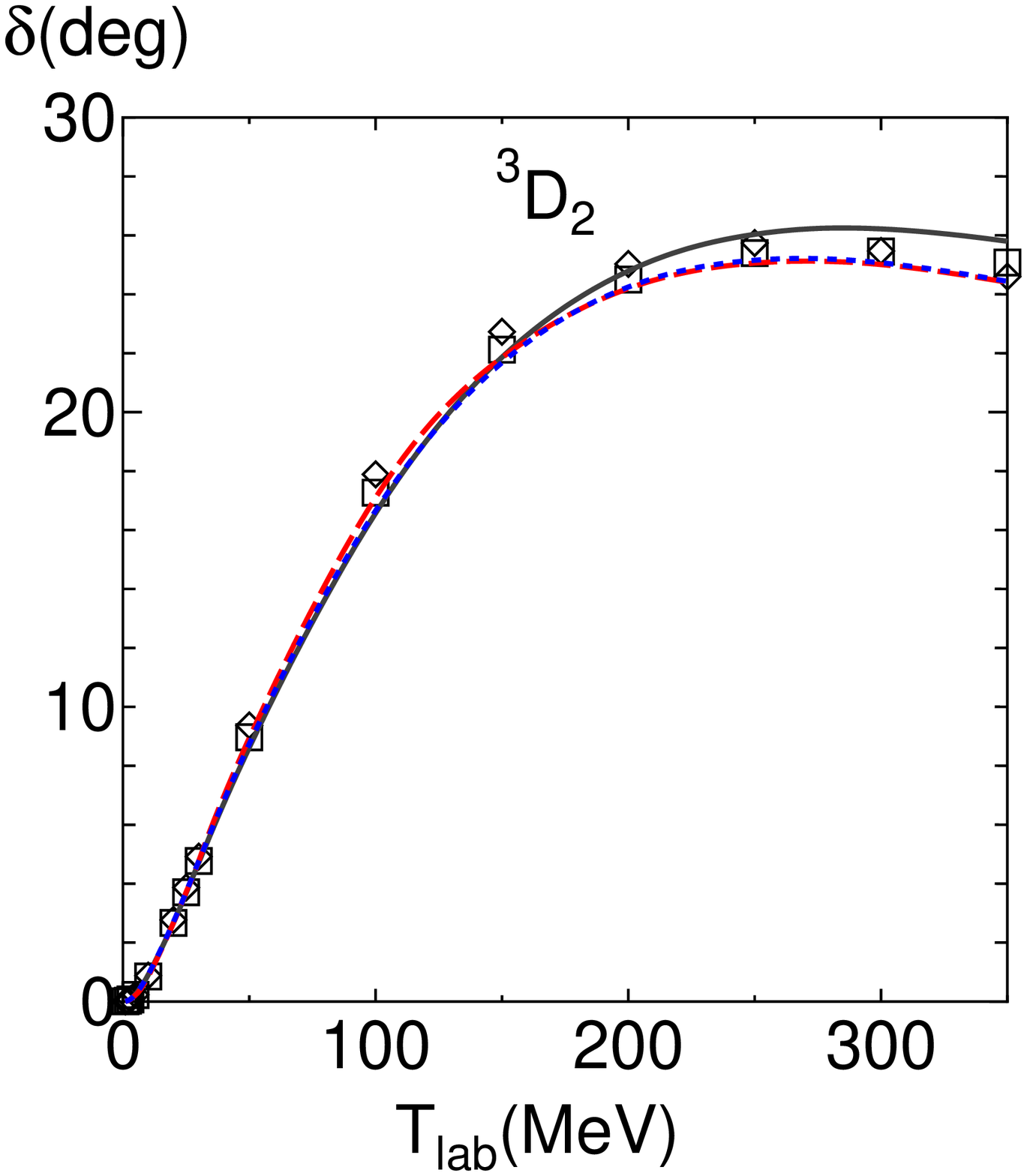}\hspace{\mys}
\includegraphics[width=\myw]{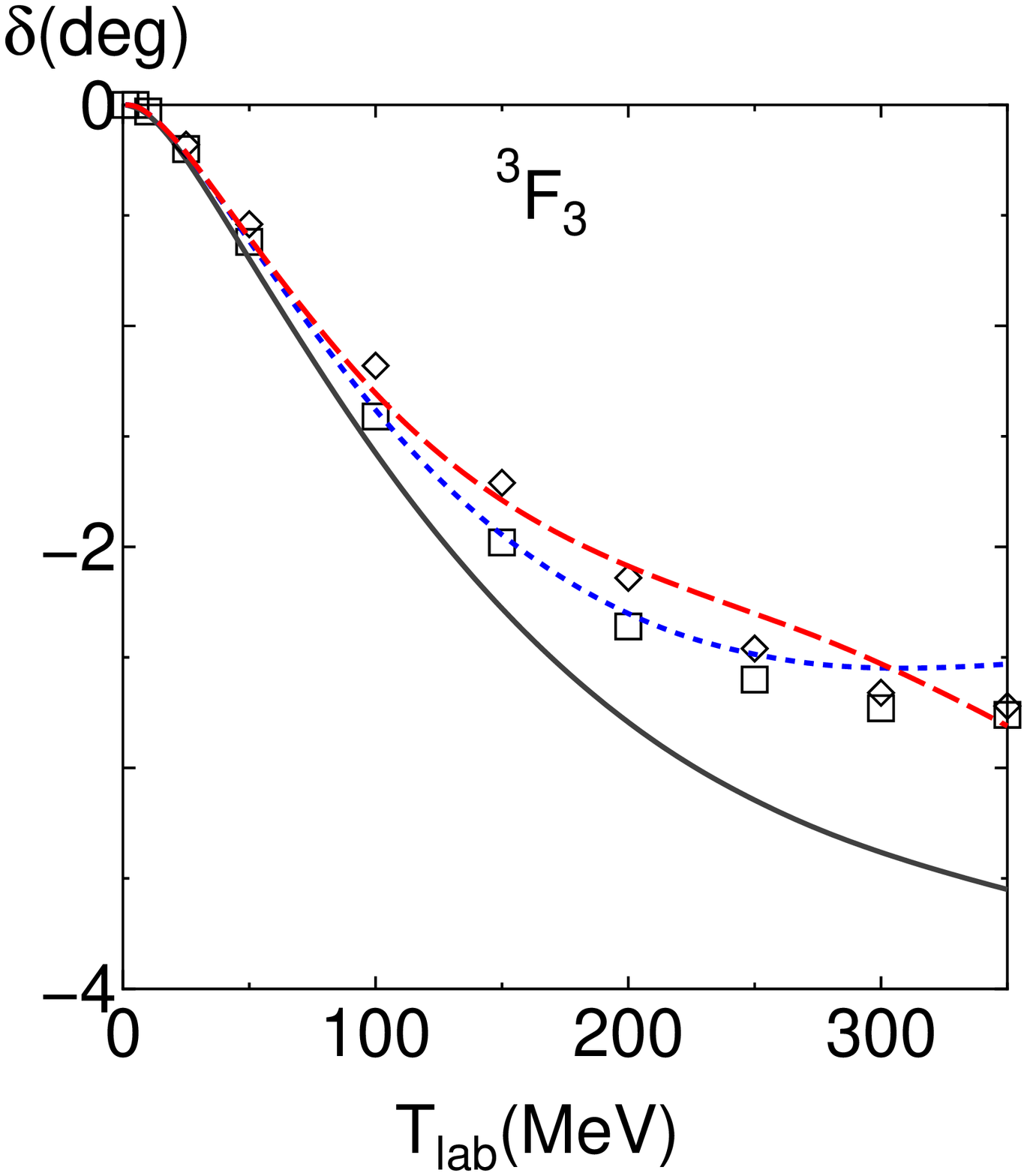}\vspace{\myv}

\vspace{10mm}
\caption{(Color online)
Phase shifts of the energy-independent Gaussian fss2 potential
up to the energy $T_\text{lab} \leq$ 350 MeV (black solid curves),
compared with the results
for the Argonne V18 \cite{AV18} (red dashed curves)
and CD-Bonn \cite{CD-Bonn} (blue short-dashed curves) potentials.
The symbols show results
from the Nijmegen multi-energy phase shift analysis (PSA) \cite{NNOL,PWA93}
and Arndt's $np$ phase shift analyses \cite{Arn83,Arn87,Arn92,Arn99}.}
\label{NNphase}
\end{figure*}

\renewcommand{\arraystretch}{1}
\begin{table*}
\caption{
Neutron-proton phase shifts and proton-proton $^1S_0$ phase shifts
(in degrees),
and $\chi^2$ values of the fss2 model
with respect to the Nijmegen PSA.}
\begin{ruledtabular}
\begin{tabular}{cdddddddddd}
$T_\text{lab}$ (MeV) &0.383&   1 &   5 &  10 &  25 &  50 & 100 & 150 & 215 & 320 \\
\hline
No.~of~data         &  144 &  68 & 103 & 290 & 352 & 572 & 399 & 676 & 756 & 954 \\
$\Delta\chi^2$      & 3034 &2100 & 274 & 710 & 919 &6804 &1409 &3913 &1464 &2637 \\
\hline
$^1S_0(pp)$ &  14.519 & 33.097 & 55.765 & 56.601 & 50.693 & 41.324 & 27.350 & 16.822 & 6.143 & -7.090 \\
$^1S_0(np)$ & 54.592 & 62.220 & 64.223 & 60.862 & 52.234 & 41.916 & 27.496 & 16.899 &  6.130 & -7.105 \\
$^3S_1$     &159.429 &147.838 &118.380 &102.881 & 80.956 & 63.014 & 43.191 & 30.495 & 18.540 &  4.737 \\
$\eps_1$    &  0.027 &  0.103 &  0.653 &  1.116 &  1.678 &  1.915 &  2.209 &  2.684 &  3.553 &  5.344 \\
$^3P_0$     &  0.048 &  0.195 &  1.747 &  3.885 &  8.546 & 11.250 &  9.038 &  4.018 & -3.147 &-14.100 \\
$^3P_1$     & -0.030 & -0.122 & -1.033 & -2.249 & -5.228 & -8.685 &-13.448 &-17.273 &-21.786 &-28.502 \\
$^1P_1$     & -0.049 & -0.193 & -1.545 & -3.183 & -6.703 &-10.267 &-14.826 &-18.392 &-22.502 &-28.421 \\
$^3P_2$     &  0.005 &  0.020 &  0.237 &  0.682 &  2.573 &  6.252 & 12.412 & 15.901 & 17.485 & 16.592 \\
$\eps_2$    & -0.000 & -0.002 & -0.057 & -0.207 & -0.824 & -1.767 & -2.854 & -3.218 & -3.211 & -2.865 \\
$^3D_1$     & -0.001 & -0.005 & -0.183 & -0.679 & -2.821 & -6.519 &-12.427 &-16.596 &-20.160 &-22.872 \\
$^3D_2$     &  0.001 &  0.006 &  0.221 &  0.842 &  3.671 &  8.825 & 17.091 & 22.260 & 22.562 & 26.213 \\
$^1D_2$     &  0.000 &  0.001 &  0.045 &  0.162 &  0.635 &  1.473 &  3.294 &  5.304 &  7.816 & 10.677 \\
$^3D_3$     &  0.000 &  0.000 &  0.001 &  0.001 &  0.008 &  0.192 &  1.287 &  2.975 &  5.361 &  8.425 \\
$\eps_3$    &  0.000 &  0.000 &  0.013 &  0.081 &  0.554 &  1.625 &  3.570 &  5.008 &  6.250 &  7.271 \\
$^3F_2$     &  0.000 &  0.000 &  0.002 &  0.013 &  0.101 &  0.319 &  0.718 &  0.978 &  1.076 &  0.595 \\
$^3F_3$     & -0.000 & -0.000 & -0.005 & -0.033 & -0.230 & -0.703 & -1.612 & -2.315 & -2.940 & -3.471 \\
$^1F_3$     & -0.000 & -0.000 & -0.011 & -0.066 & -0.424 & -1.156 & -2.344 & -3.194 & -4.028 & -5.094 \\
$^3F_4$     & -0.000 &  0.000 &  0.000 &  0.001 &  0.017 &  0.087 &  0.366 &  0.816 &  1.642 &  3.455 \\
$\eps_4$    & -0.000 & -0.000 & -0.000 & -0.004 & -0.047 & -0.191 & -0.540 & -0.866 & -1.223 & -1.642 \\
$^3G_3$     & -0.000 & -0.000 & -0.000 & -0.004 & -0.054 & -0.264 & -0.982 & -1.882 & -3.109 & -4.889 \\
$^3G_4$     &  0.000 &  0.000 &  0.001 &  0.014 &  0.171 &  0.726 &  2.191 &  3.673 &  5.423 &  7.716 \\
$^1G_4$     &  0.000 &  0.000 &  0.000 &  0.003 &  0.038 &  0.147 &  0.386 &  0.606 &  0.886 &  1.413 \\
$^3G_5$     &  0.000 & -0.000 & -0.000 & -0.000 & -0.009 & -0.052 & -0.189 & -0.316 & -0.399 & -0.279 \\
$\eps_5$    &  0.000 &  0.000 &  0.000 &  0.002 &  0.037 &  0.206 &  0.733 &  1.300 &  1.983 &  2.917
\end{tabular}
\end{ruledtabular}
\label{tab:nnphase}
\end{table*}

\renewcommand{\arraystretch}{1.0}
\begin{table*}[htbp]
\begin{ruledtabular}
\caption{
Comparison of the deuteron properties with the predictions using
fss2, Bonn~C, CD-Bonn, and experimental data.}
\begin{tabular}{l|lllll}
 & fss2-Gauss
 & fss2 (isospin) \cite{fss2NN}
 & Bonn C \cite{BonnC}
 & CD-Bonn \cite{CD-Bonn} & Expt. \\
\hline
$\eps_d$ (MeV)     & 2.2206 & 2.2250 & fitted & fitted
                   & 2.224644$\pm$0.000046 \cite{NPB.216.277.(1983)}\\
$R_\text{rms}$ (fm)& 1.961  & 1.960  & 1.968  & 1.966
                   & 1.971$\pm$0.006 \cite{PRL.70.2261.(1993),PRA.49.2255.(1994),PRC.51.1127.(1995)}\\
$Q_d$ (fm$^2$)     & 0.270  & 0.270  & 0.281  & 0.270
                   & 0.2859$\pm$0.0003 \cite{NPA.405.497.(1983),PRA.20.381.(1979)}\\
$\eta=A_D/A_S$     & 0.0252 & 0.0253 & 0.0266 & 0.0256
                   & 0.0256$\pm$0.0004 \cite{PRC.41.898.(1990)}\\
$P_D$ (\%)         & 5.52   & 5.49   & 5.60   & 4.85
\label{D-properties}
\label{force}
\end{tabular}
\end{ruledtabular}
\end{table*}

\begin{figure}[htbp]
\includegraphics[width=0.9\hsize,trim=0 0 0 0]{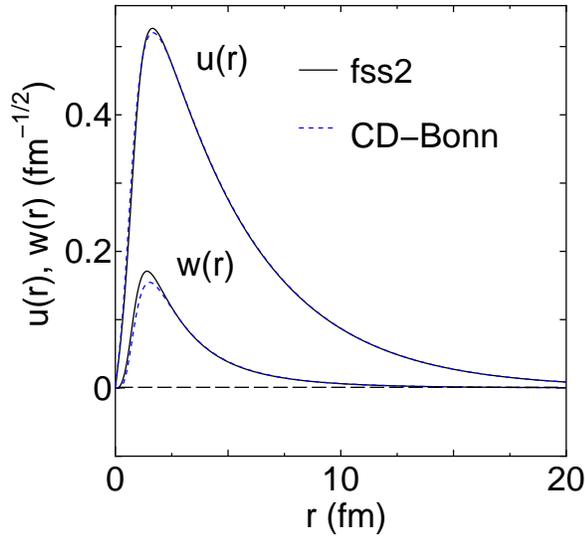}
\vspace{-3mm}
\caption{(Color online)
The deuteron wave functions
$u(r)=u_{0\alpha}(r)$ and $w(r)=u_{2\alpha}(r)$
predicted by the energy-independent Gaussian potential based on fss2
(black solid curve),
compared with CD-Bonn (blue dashed) \cite{CD-Bonn}.}
\label{deuteron1}
\end{figure}

\begin{figure}[t]
\includegraphics[width=0.9\hsize,trim=0 0 0 0]{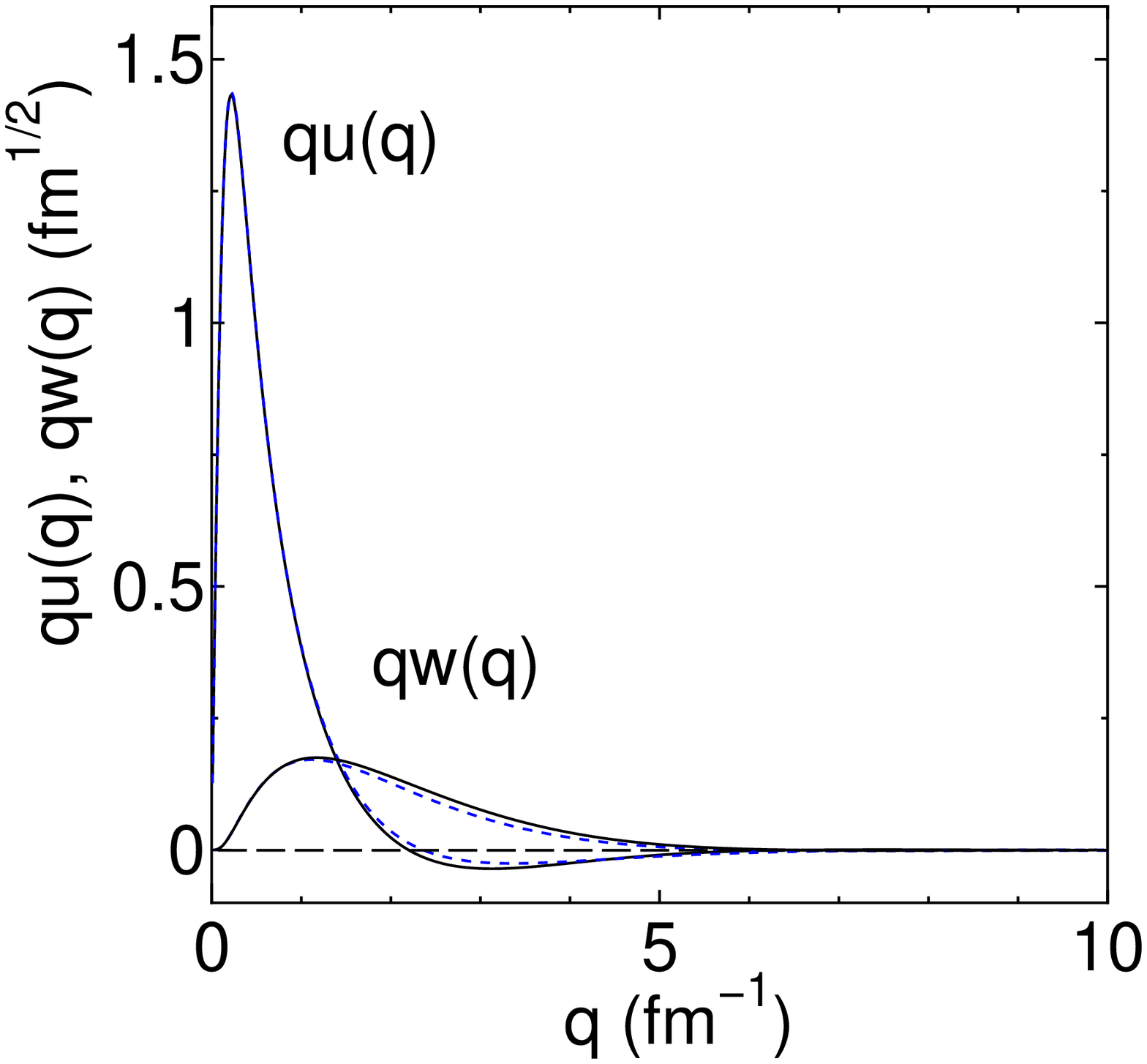}
\vspace{-3mm}
\caption{(Color online)
As Fig.~\ref{deuteron1},
but for the deuteron wave functions in momentum space
$qu(q)=qf_{0\alpha}(q)$ and
$qw(q)=qf_{2\alpha}(q)$.}
\label{deuteron2}.
\end{figure}

In this study,
we use a Gaussian representation of the fss2 potential
for numerical simplicity \cite{Fuk09}.
This representation conserves the non-local feature of the potential accurately
and reproduces the original phase shifts within an accuracy better than
$0.1^\circ$ for almost all energies and partial waves.
The energy dependence of the Gaussian-represented potential
is eliminated by the above-mentioned method.
The phase shifts for the energy-independent version
of fss2 are shown in Fig.~\ref{NNphase}.
One can find some deficiencies in the $^3P_2$ and $^3F_{2,3}$ phase shifts.
As already pointed in Ref.~\cite{fss2NN},
this is probably related to the problem of the balance
of central and $LS$ force in the short-range region.
Appreciable deviations appear also in the $^3D_1$ and $^3D_3$ channels
in the higher-energy region.
This implies that improvements of the tensor force are desirable
in future refinements of the interaction.

Examining $\chi^2$ with respect to the PSA is a good test
to see how strict the interaction describes the observables.
In this study,
we worked in the isospin basis and with the cut-off Coulomb force
\be
 V_C(k) = \frac{1-\cos(kR_\text{cut})}{k^2} \:,
\label{eq:coulomb}
\ee
and extracted the nuclear phase shifts by using the
Vincent-Phatak method \cite{PRC.10.391.(1974)}.
This method gives stable phase shifts with respect to the change of
$R_\text{cut}$ \cite{pdscat},
which we take as $R_\text{cut}=10$ fm.

Another important factor in considering the $pp$ phase shifts
is the charge-independence breaking (CIB).
The CIB effect is taken into account by a reduction factor for the
coupling constant of the scalar-singlet meson,
which is determined to minimize $\chi^2$ \cite{fss2NN,QMalpha}.
Our value of the reduction factor is 0.9932,
which is quite close to 0.9934 used in Ref.~\cite{QMalpha}.
We report in Tab.~\ref{tab:nnphase} the neutron-proton ($np$)
and the $^1S_0$ proton-proton ($pp$) phase shifts with their
$\Delta\chi^2$ values with respect to the Nijmegen PSA \cite{PWA93,Tom}.
The Gaussian fss2 potential gives $\chi^2/N_\text{data}=6.34$.
Although the phase shifts are overall well reproduced,
as we saw in Fig.~\ref{NNphase},
we have larger $\chi^2$ values mainly due to the $^3P_2$, $^1D_2$, and $^3D_3$
partial waves in the high-energy region.
Moreover, the constraint for the $^1S_0$ $pp$ phase shifts in the
low-energy region is very severe.
Those phase shifts given by the Nijmegen PSA are
$14.609^\circ$ and $32.688^\circ$
at $E_p=0.383$ MeV and $1$ MeV, respectively \cite{Tom}.
A more developed treatment is desirable,
because the difference from the Nijmegen PSA should be much less
than $0.1^\circ$.

When the deuteron properties were determined in Appendix~B of Ref.~\cite{fss2NN},
the authors first solved Eq.~(\ref{ene-dep}) for $\chi(\rv)$
and then obtained the renormalized relative wave function
$\Psi(\rv)=N^{1/2}\chi(\rv)$.
For the renormalized potential,
we can directly solve Eq.~(\ref{ene-indep}) for $\Psi(\rv)$.
Because the procedure to obtain the potential and the deuteron wave function
is different,
it is necessary to reexamine the deuteron properties.
The detailed prescription is explained in Appendix~A.

Figures~\ref{deuteron1} and \ref{deuteron2} show the deuteron wave functions
by the energy-independent fss2 potential in coordinate and momentum space,
respectively.
The wave functions are indistinguishable from those of
Ref.~\cite{fss2NN}.
It can also be seen from Table \ref{D-properties} that
the renormalization of the kernel does not change the deuteron properties.
Our calculation slightly underpredicts the quadrupole moment,
as other potentials do.
According to Refs.~\cite{All78,Koh83},
the correction due to the meson-exchange current is typically
about 0.01 $\text{fm}^2$.
However, even taking into account this correction,
there is still a small discrepancy.

\begin{figure}[tbp]
\includegraphics[width=80mm]{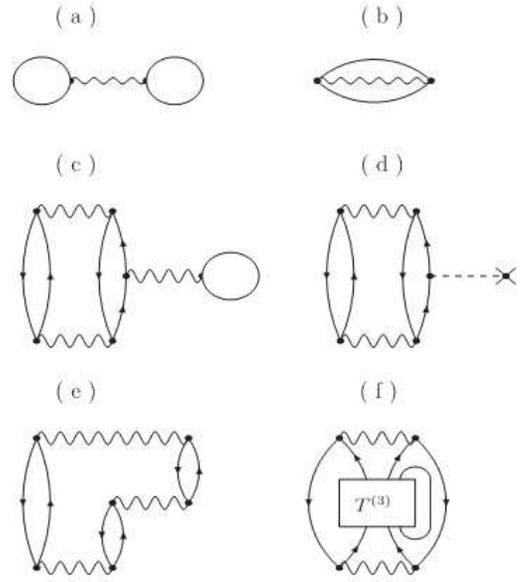}
\caption{
Different Goldstone diagrams contributing to the nuclear matter EOS.
Diagrams (a) and (b) correspond to the BHF calculation.
The sum of the other diagrams gives the three-hole-line contribution.
For more details, please see the text.}
\label{fig:3HL}
\end{figure}

\section{Nuclear matter EOS within the Bethe-Brueckner-Goldstone approach}
\label{s:bhf}

\subsection{Sketch of the approach}

The basis of the BHF calculation is the Bethe-Goldstone equation
for the $G$ matrix,
\bea
 && \lla 12|G(\omega)|34 \rra_A = \lla 12|V|34 \rra_A
\nonumber\\ && \hspace{5mm}
 + \sum_{5,6} \lla 12|V|56 \rra \frac{Q_{56}}{\omega-e_5-e_6}
 \lla 56|G(\omega)|34 \rra_A \:,
\label{eq:BG-eq}
\eea
where the multi-indices $1,2,\ldots$ include the
momentum and the spin-isospin variables of the particles,
$|12\rangle_A \equiv |12\rangle-|21 \rangle$,
$V$ is the bare nucleon-nucleon interaction,
$\omega$ is the starting energy,
and $Q_{56} \equiv \Theta(k_5-k_F^{(5)}) \Theta(k_6-k_F^{(6)})$
is the two-particle Pauli operator,
where $k_F^{(i)}$ are the Fermi momenta of the nucleons.
In this study, we use the angle-averaged form of the Pauli operator
and the two-particle intermediate energy $e_5+e_6$,
in order to avoid the complex coupling of the angular-momentum quantum numbers
\cite{angleav}.

\def\myc#1{\multicolumn{1}{c}{$#1$}}
\begin{table*}[t]
\caption{
Energies per nucleon (in MeV) of the various
THL contributions to the SNM EOS for different
Fermi momenta $k_F$ in fm$^{-1}$.
The continuous choice is adopted.
The baryon density $\rho=2k_F^3/(3\pi^2)$ in fm$^{-3}$ is also shown.
$T+E_2$ corresponds to the BHF calculations,
$B$ is the bubble diagram,
$BU$ is the $U$-insertion diagram,
$R$ is the ring diagram,
$H$ indicates the higher-order diagrams,
and $E_3$ is the total THL contribution.
}
\begin{ruledtabular}
\begin{tabular}{ddddddddd}
 \myc{k_F}&\myc{\rho}&\myc{T+E_2}&\myc{B}&\myc{BU}&\myc{R}&\myc{H\times10^3}&\myc{E_3}&\myc{T+E_2+E_3} \\
\hline
  1.1 & 0.090 & -17.07 & -7.14 & 11.02 & -0.77 & 186.8 &  3.30 & -13.77 \\
  1.2 & 0.117 & -19.58 & -6.17 & 11.46 & -1.27 & 159.3 &  4.22 & -15.36 \\
  1.3 & 0.148 & -21.98 & -4.37 & 11.68 & -1.63 & 113.8 &  5.79 & -16.19 \\
  1.4 & 0.185 & -24.21 & -2.33 & 12.48 & -1.87 & 77.22 &  8.35 & -15.86 \\
  1.5 & 0.228 & -26.91 &  0.80 & 12.93 & -1.99 & 48.35 & 11.79 & -14.30 \\
  1.6 & 0.277 & -27.41 &  4.93 & 13.91 & -2.13 & 26.58 & 16.73 & -10.67 \\
  1.7 & 0.332 & -27.90 & 10.76 & 14.83 & -2.05 & 10.88 & 23.56 &  -4.34 \\
  1.8 & 0.394 & -27.33 & 18.91 & 15.61 & -1.93 &  1.19 & 32.59 &   5.27 \\
  1.9 & 0.463 & -25.57 & 29.85 & 16.21 & -1.75 & -3.75 & 44.31 &  18.74 \\
  2.0 & 0.540 & -22.25 & 43.71 & 15.09 & -1.42 &  2.49 & 57.39 &  35.14 \\
  2.1 & 0.626 & -17.08 & 61.42 & 13.52 & -1.29 &  0.80 & 73.66 &  56.57 \\
  2.2 & 0.719 &  -9.73 & 83.58 & 10.47 & -1.19 & -0.11 & 92.86 &  83.13 \\
  2.3 & 0.822 &  -0.01 &111.12 &  6.15 & -1.11 & -0.21 &116.16 & 116.16 \\
\end{tabular}
\end{ruledtabular}
\label{SNM-cont}
\end{table*}

\begin{table*}
\vspace{-2mm}
\caption{As Table~\ref{SNM-cont},
but with the gap choice for the s.p.~potential.}
\begin{ruledtabular}
\begin{tabular}{dddddddd}
 \myc{k_F}&\myc{\rho}&\myc{T+E_2}&\myc{B}&\myc{R}&\myc{H\times10^3}&\myc{E_3}&\myc{T+E_2+E_3} \\
\hline
1.1 & 0.090 & -11.54 & -0.48 & -1.00 & 65.07 & -1.42 & -12.95 \\
1.2 & 0.117 & -13.45 &  0.08 & -1.09 & 49.27 & -0.96 & -14.41 \\
1.3 & 0.148 & -15.26 &  1.02 & -1.18 & 33.16 & -0.13 & -15.39 \\
1.4 & 0.185 & -16.86 &  2.48 & -1.22 & 20.72 &  1.28 & -15.59 \\
1.5 & 0.228 & -18.12 &  4.64 & -1.25 & 11.66 &  3.40 & -14.72 \\
1.6 & 0.277 & -18.88 &  7.82 & -1.29 &  5.45 &  6.54 & -12.34 \\
1.7 & 0.332 & -19.00 & 12.40 & -1.26 &  0.99 & 11.14 &  -7.85 \\
1.8 & 0.394 & -18.31 & 18.96 & -1.24 & -1.99 & 17.72 &  -0.59 \\
1.9 & 0.463 & -16.65 & 28.03 & -1.19 & -3.96 & 26.84 &  10.19 \\
2.0 & 0.540 & -13.90 & 39.71 & -1.09 & -1.80 & 38.62 &  24.73 \\
2.1 & 0.626 &  -9.90 & 55.18 & -1.06 & -1.83 & 54.12 &  44.22 \\
2.2 & 0.719 &  -4.53 & 75.75 & -1.95 & -1.76 & 74.71 &  70.17 \\
2.3 & 0.822 &   2.29 &102.63 & -1.06 & -1.56 &101.57 & 103.85 \\
\end{tabular}
\end{ruledtabular}
\label{SNM-gap}
\end{table*}

The single-particle (s.p.) energy is defined by
$e_i = k_i^2/2M_N+U_i(k_i)$,
where $M_N$ is the nucleon mass.
The auxiliary s.p.~potentials $U_i(k_i)$
are self-consistently determined by the on-shell $G$ matrix elements,
with Eq.~(\ref{eq:BG-eq}),
\be
 U_i(k_i) = \sum_{j<k_F} \lla ij \left| G(e_i+e_j) \right| ij \rra_A \:.
\label{eq:sppot}
\ee
The s.p.~potential $U(k)$ can be chosen in various ways.
Our investigations will be carried out for two somewhat opposite choices:
the continuous choice and the gap (or standard) choice.
In the continuous choice, Eq.~(\ref{eq:sppot}) is solved for all $k$,
while $U(k)=0$ is assumed for $k>k_F$ in the gap choice.
The detailed procedure for BHF calculations is presented
in Refs.~\cite{book} and \cite{Bal91}.

In this study,
we calculate the binding energy per particle for symmetric nuclear matter (SNM)
$E/A(\rho,x_p=0.5)$ and pure neutron matter (PNM) $E/A(\rho,x_p=0)$,
where $x_p=\rho_p/\rho$ is the proton fraction.
In these cases the energy per particle from the two-hole-line contribution
is given by
\be
 \left(\frac{E}{A}\right)_2 =
 \frac{3}{5} \frac{k_F^2}{2M_N} + \frac{1}{2\rho} \sum_{k<k_F}U(k) \:.
\ee
At a given baryonic density $\rho$,
we will approximate $E/A$ for asymmetric nuclear matter
by the parabolic approximation,
\be
 \frac{E}{A}(\rho,x_p) =
 (1-\beta) \frac{E}{A}\Bigg|_\text{SNM}\hspace{-1.6em}(\rho)
   + \beta \frac{E}{A}\Bigg|_\text{PNM}\hspace{-1.6em}(\rho) \:,
\label{e:parab}
\ee
where $\beta=(1-2x_p)^2$,
which has been verified to be a good approximation within the BHF approach
\cite{Bom91}.

\begin{table*}
\caption{As Table~\ref{SNM-cont}, but for PNM.}
\begin{ruledtabular}
\begin{tabular}{ddddddddd}
 \myc{k_F}&\myc{\rho}&\myc{T+E_2}&\myc{B}&\myc{BU}&\myc{R}&\myc{H\times10^3}&\myc{E_3}&\myc{T+E_2+E_3} \\
\hline
  1.0 & 0.034 &  5.18 & -0.84 &   1.17 & -0.01 &   2.63 &  0.33 &   5.50 \\
  1.1 & 0.045 &  5.96 & -0.85 &   1.38 & -0.11 &  -0.04 &  0.42 &   6.38 \\
  1.2 & 0.058 &  6.83 & -0.76 &   1.53 & -0.15 &  -3.74 &  0.62 &   7.45 \\
  1.3 & 0.074 &  7.66 & -0.63 &   1.84 & -0.15 &  -7.02 &  1.04 &   8.71 \\
  1.4 & 0.093 &  8.55 & -0.38 &   2.14 & -0.15 &  -9.34 &  1.60 &  10.15 \\
  1.5 & 0.114 &  9.56 &  0.07 &   2.41 & -0.13 & -10.59 &  2.34 &  11.90 \\
  1.6 & 0.138 & 10.71 &  0.74 &   2.81 & -0.10 & -11.32 &  3.44 &  14.14 \\
  1.7 & 0.166 & 12.10 &  1.79 &   3.25 & -0.09 & -11.17 &  4.94 &  17.04 \\
  1.8 & 0.197 & 13.87 &  3.40 &   3.66 & -0.09 & -10.39 &  6.96 &  20.83 \\
  1.9 & 0.232 & 16.05 &  5.69 &   4.11 & -0.09 &  -8.81 &  9.70 &  25.75 \\
  2.0 & 0.270 & 18.76 &  8.74 &   4.30 & -0.05 &  -5.54 & 12.98 &  31.74 \\
  2.1 & 0.313 & 22.23 & 12.81 &   4.39 & -0.07 &  -4.17 & 17.13 &  39.36 \\
  2.2 & 0.360 & 26.61 & 18.92 &   4.14 & -0.08 &  -2.99 & 22.97 &  49.59 \\
  2.3 & 0.411 & 32.02 & 24.66 &   3.66 & -0.10 &  -2.09 & 28.22 &  60.23 \\
  2.4 & 0.467 & 38.53 & 33.09 &   2.76 & -0.10 &  -1.33 & 35.74 &  74.28 \\
  2.5 & 0.528 & 46.42 & 43.30 &   1.18 & -0.11 &  -0.75 & 44.36 &  90.79 \\
  2.6 & 0.594 & 55.86 & 55.46 &  -1.08 & -0.13 &  -0.39 & 54.25 & 110.11 \\
  2.7 & 0.665 & 66.93 & 70.50 &  -4.14 & -0.14 &  -0.15 & 66.22 & 133.15 \\
  2.8 & 0.741 & 79.78 & 89.54 &  -8.23 & -0.15 &  -0.09 & 81.16 & 160.94 \\
  2.9 & 0.824 & 94.52 &110.30 & -13.50 & -0.27 &   0.42 & 96.53 & 191.05 \\
\end{tabular}
\end{ruledtabular}
\label{PNM-cont}
\end{table*}

\begin{table*}
\caption{As Table~\ref{PNM-cont},
but with the gap choice for the s.p.~potential.}
\begin{ruledtabular}
\begin{tabular}{ddddddddd}
 \myc{k_F}&\myc{\rho}&\myc{T+E_2}&\myc{B}&\myc{R}&\myc{H\times10^3}&\myc{E_3}&\myc{T+E_2+E_3} \\
\hline
1.0 & 0.034 &  5.79 & -0.23 & -0.05 &  1.03 & -0.28 &   5.51 \\
1.1 & 0.045 &  6.70 & -0.18 & -0.08 & -0.82 & -0.26 &   6.43 \\
1.2 & 0.058 &  7.63 & -0.09 & -0.10 & -2.68 & -0.18 &   6.51 \\
1.3 & 0.074 &  8.62 &  0.07 & -0.09 & -4.21 & -0.03 &   8.59 \\
1.4 & 0.093 &  9.69 &  0.32 & -0.09 & -5.10 &  0.23 &   9.91 \\
1.5 & 0.114 & 10.88 &  0.71 & -0.07 & -5.42 &  0.63 &  11.50 \\
1.6 & 0.138 & 12.25 &  1.30 & -0.06 & -5.56 &  1.24 &  13.49 \\
1.7 & 0.166 & 13.88 &  2.19 & -0.05 & -5.45 &  2.14 &  16.02 \\
1.8 & 0.197 & 15.85 &  3.53 & -0.05 & -5.16 &  3.47 &  19.32 \\
1.9 & 0.232 & 18.24 &  5.43 & -0.06 & -4.60 &  5.37 &  23.61 \\
2.0 & 0.270 & 21.15 &  7.95 & -0.05 & -3.11 &  7.90 &  29.06 \\
2.1 & 0.313 & 24.69 & 11.34 & -0.06 & -2.41 & 11.28 &  35.97 \\
2.2 & 0.360 & 28.96 & 15.89 & -0.08 & -1.79 & 15.19 &  44.77 \\
2.3 & 0.411 & 34.40 & 21.76 & -0.09 & -1.26 & 21.67 &  55.71 \\
2.4 & 0.467 & 40.02 & 29.50 & -0.10 & -0.83 & 29.40 &  69.43 \\
2.5 & 0.528 & 47.00 & 39.64 & -0.12 & -0.47 & 39.53 &  86.52 \\
2.6 & 0.594 & 55.02 & 52.63 & -0.14 & -0.24 & 52.49 & 107.51 \\
2.7 & 0.665 & 64.13 & 69.74 & -0.16 & -0.06 & 69.58 & 133.71 \\
2.8 & 0.741 & 74.37 & 93.04 & -0.19 & -0.05 & 92.85 & 167.22 \\
2.9 & 0.824 & 85.70 &123.18 & -0.21 &  0.08 &122.97 & 208.67 \\
\end{tabular}
\end{ruledtabular}
\label{PNM-gap}
\end{table*}

The two-hole-line and three-hole-line (THL) diagrams are depicted
in Fig.~\ref{fig:3HL},
where the wavy line denotes the $G$ matrix.
Figures~\ref{fig:3HL}(a) and \ref{fig:3HL}(b) are the above-mentioned
BHF direct (Hartree) and exchange (Fock) diagram, respectively.
As for the THL calculations,
we closely follow the method described in detail in Ref.~\cite{Day}.
The full scattering process of three particles that
are virtually excited above the Fermi sphere can be calculated
by solving the Bethe-Faddeev equation
for the in-medium three-body scattering matrix $T^{(3)}$,
as depicted in Fig.~\ref{fig:3HL}(f) \cite{book,Day,Raj67}.
For computational convenience,
the lowest-order contribution in the $G$ matrix,
namely diagram \ref{fig:3HL}(c),
is calculated separately.
Figure~\ref{fig:3HL}(c) is known as the ``bubble" diagram,
and Fig.~\ref{fig:3HL}(d) is the corresponding $U$-insertion diagram.
Note that $U$-insertion diagram vanishes in the gap choice,
because $U(k)=0$ is assumed for $k>k_F$.
Figure~\ref{fig:3HL}(e) is the ``ring" diagram,
which is responsible for long-range correlations in nuclear matter.
An indication of convergence of the BBG expansion
is the possible small size of the THL
contribution with respect to the two-hole-line contribution.

\subsection{Numerical results}

\begin{table}[tbp]
\caption{
The fitted coefficients in Eq.~(\ref{eq:Polyfit}) for SNM and PNM
and continuous (C) or gap (G) choice,
together with those obtained for other interactions.}
\begin{ruledtabular}
\begin{tabular}{ldddd}
            &    a   &   b   &  c   &  d   \\
\hline
SNM,C       & -515.4 & 692.6 & 1.30 &  2.4 \\
SNM,G       & -113.6 & 296.4 & 1.91 & -6.2 \\
PNM,C       & -103.5 & 355.3 & 1.48 &  8.3 \\
PNM,G       &   60.2 & 252.3 & 2.50 &  3.4 \\
\hline
SNM,APR     & -101.5 & 333.9 & 2.14 & -4.8 \\
SNM,DBHF    & -422.8 & 711.3 & 1.56 &  7.3 \\
SNM,V18+TBF & -123.2 & 407.9 & 2.38 &  0.  \\
SNM,V18+UIX & -137.0 & 308.0 & 1.82 & -5.0 \\
PNM,APR     &   76.1 & 256.5 & 2.71 &  3.6 \\
PNM,DBHF    & -230.3 & 715.5 & 1.58 &  9.5 \\
PNM,V18+TBF &   55.9 & 532.3 & 2.68 &  0.  \\
PNM,V18+UIX &   11.0 & 309.0 & 1.95 &  6.0 \\
\end{tabular}
\end{ruledtabular}
\end{table}

We report in Tables~\ref{SNM-cont}-\ref{PNM-gap}
the contributions of each diagram to the EOS
for SNM and PNM in the continuous and gap choice, respectively.
The slightly different results with respect to Ref.~\cite{PRL2014}
are due to the more refined momentum grid we used in the present calculations.
This was relevant at higher density in order to obtain convergence in the BHF
iteration procedure.
We divide the space of relative momentum $q$ into the two domains,
$[0,a]$ and $[a,\infty]$
and apply the Gauss-Legendre quadrature to each part
in solving the Bethe-Goldstone equation.
The mapping $q=a+\tan{[(1+x)\pi/4]}$,
where $x$ are the nodes of the Gauss-Legendre quadrature,
is used for the second part as in Ref.~\cite{LS-RGM}.
In this study, $a=6\;\text{fm}$ is adopted and we take 70 points
in the first section and 30 points in the second section.
All nucleon-nucleon channels up to the total angular momentum $J=8$
were considered.
In each iteration,
the s.p.~potential was calculated self-consistently up to
$k_\text{max}=7.5~\text{fm}^{-1}$
with a grid step of $0.1\;\text{fm}^{-1}$.
After 30 iterations, a convergence within few keV was reached
in all calculations at $k_F \leq 3.0\;\text{fm}^{-1}$.

After adding the THL contributions,
we fitted the calculated EOS,
both for SNM and PNM, by an analytic form with four parameters
\be
 \frac{E}{A}(\rho) = a\rho + b\rho^c + d \:.
\label{eq:Polyfit}
\ee
The values of the fitted parameters are listed in Table VII.
These fitted EOS are valid for not too low density
$\rho\gtrsim0.1\;\text{fm}^{-3}$
and must not be extrapolated to zero density.
In the continuous choice around saturation the
analytic form is very close to the calculated points,
and one can extract the saturation energy
$(E/A)_0=-16.3\;\text{MeV}$
at $\rho_0=0.157\;\text{fm}^{-3}$
and an incompressibility $K=219\;\text{MeV}$,
while with the gap choice one finds
$(E/A)_0=-15.6\;\text{MeV}$,
$\rho_0=0.170\;\text{fm}^{-3}$,
and $K=185\;\text{MeV}$.
This indicates the uncertainty of the EOS around saturation.
At higher density the fit is less precise,
but the deviation does not exceed 1.2 MeV even at the highest density
in the continuous choice,
which is more than enough for NS calculations.
In the gap choice,
it is not so easy to describe the EOS by one single analytic expression
from very low to high density as in the continuous choice,
but the reported values of the parameters give a good fit at high density,
which is useful for the NS study.

\begin{figure}
\includegraphics[width=0.9\hsize]{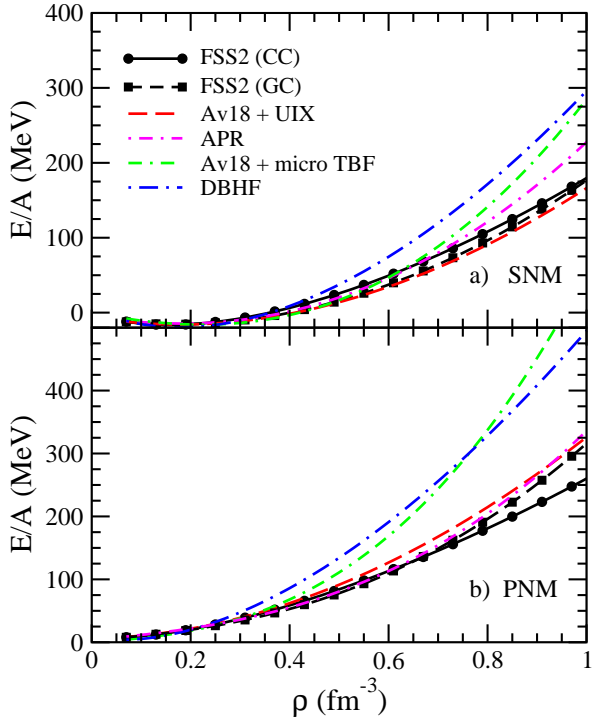}
\caption{(Color online)
The energy per particle of SNM (upper panel) and PNM (lower panel)
for several EOS as a function of baryon density.
The curves FSS2 (CC) and FSS2 (GC) show the calculated fss2 EOS
in continuous and gap choice, respectively.
The curve $Av_{18}$+UIX refers to the BHF calculation
using Argonne V18 plus Urbana IX TBF from Ref.~\cite{Urb,Tar},
the curve APR is the variational calculation from Ref.~\cite{APR},
the curve $Av_{18}$+micro TBF is the BHF calculation from Ref.~\cite{Umb,Hans},
and the curve DBHF is the relativistic DBHF calculation from Ref.~\cite{DBHF}.}
\label{fig:EOS}
\end{figure}

The fss2 EOS for SNM and PNM are reported in Fig.~\ref{fig:EOS},
in comparison with the corresponding EOS from some other approaches.
The latter have been selected from the ones that are able to reproduce
the saturation point within the phenomenological uncertainty.
The comparison of the different EOS for SNM shows a
substantial agreement up to about 0.5 fm$^{-3}$,
while at higher density both the variational calculation (APR)
of Ref.~\citep{APR}
and the relativistic DBHF calculation
of Ref.~\cite{DBHF} indicate a stiffer trend.
As for non-relativistic BHF calculations,
also the EOS obtained with the ``microscopic'' TBF of Refs.~\cite{Umb,Hans}
is appreciably stiffer in that density region,
while the BHF with the Urbana model \cite{Urb,Tar} for the TBF
produces an EOS in substantial agreement with the fss2 EOS.
A similar trend is present in PNM.

In Fig.~\ref{fig:EOS} the EOS is reported in the density range relevant
for NS calculations.
From Tables~\ref{SNM-cont}-\ref{PNM-gap}
one can notice that at the highest densities
the THL contribution can be larger than the two-hole-line one.
This can make the convergence of the BBG expansion at least questionable.
However, it turns out that the two-hole-line interaction part
is quite limited because of the strong compensation between negative and
positive channel contributions,
and actually it is decreasing in the highest density region.
On the contrary, the dominant contribution to the THL
interaction term is coming from the repulsive ``bubble" diagram.
This consideration can suggest that the convergence might still be present,
but of course it cannot give a strong argument in support of it.

However, in addition to that,
one can see from Tables~\ref{SNM-cont}-\ref{PNM-gap} that
the three-body scattering processes,
described by the scattering matrix $T^{(3)}$ (column ``$H$"),
give a negligible contribution.
It can then be expected that the four-body scattering processes
will also be negligibly small.
The fourth-order diagrams apart from the four-body scattering processes,
have been estimated in Ref.~\cite{Day4} up to about three times saturation
density for the Reid soft-core NN interaction \cite{Reid68}
and found to be quite small.
We will assume that this is also true in our case,
even for higher density,
at least approximately enough for NS calculations.

As a check of this assumption we will confront the EOS with known
phenomenological constraints.
The higher density part of the EOS,
needed for NS calculations,
can be seen as an extrapolation from the lower one,
which can be validated from the comparison with astrophysical observations
and laboratory experiments on heavy-ion collisions.
It is clear that the main theoretical uncertainty on the EOS is indeed
coming from the many-body calculations at higher density.
Unfortunately it is difficult to get a quantitative estimate
of this uncertainty without a firm limit
on the higher-order contributions beyond the THL.

\subsection{Comparison with phenomenology}

\begin{figure}
\includegraphics[width=1.\hsize,clip=true]{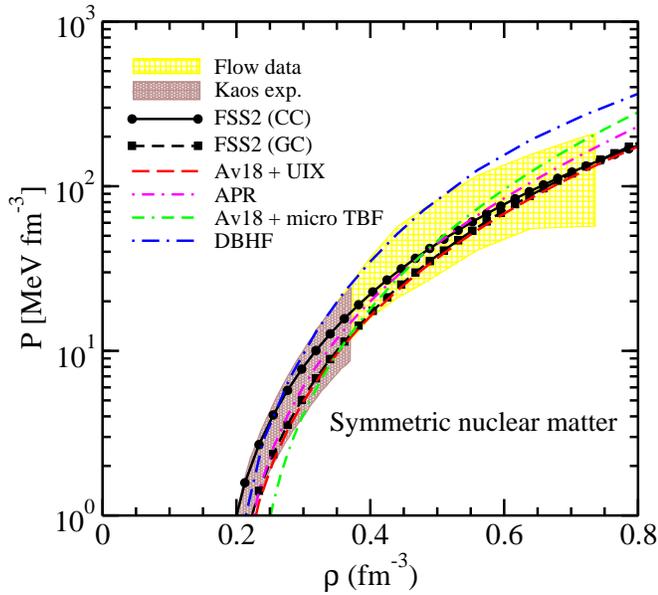}
\caption{(Color online)
The pressure of symmetric matter for several EOS.
The larger (yellow) and smaller (violet) bands
represent the phenomenological constraints
from experimental data.
See text for details.}
\label{fig:p_sym}
\end{figure}

We will now confront the EOS with a set of phenomenological constraints
in order to assess its reliability.
Possible tests of the EOS have been devised from experiments on HIC,
which have been performed in the last two decades at energies
ranging from few tens to several hundreds MeV per nucleon.
It can be expected in fact that in HIC at large enough energy
nuclear matter is compressed and that
the two partners of the collisions produce flows of matter.
In principle the dynamics of the collisions should be connected
with the properties of the nuclear medium EOS and its viscosity.
In the so-called ``multifragmentation" regime,
after the collision numerous nucleons and fragments of different sizes
are emitted,
and the transverse flow,
which is strongly affected by the matter compression during the collision,
can be measured.
Based on numerical simulations,
it was proposed in Ref.~\cite{DL}
that any reasonable EOS for SNM
should pass through a phenomenological region in the pressure vs.~density plane.

The plot is reproduced in Fig.~\ref{fig:p_sym},
where a comparison with the same microscopic calculations is made.
The larger (yellow) dashed box represents the results of the numerical
simulations of the experimental data discussed in Ref.~\cite{DL},
and the smaller (violet) one represents the constraints from the
experimental data on kaon production \cite{Fuchs}.
The fss2 EOS
is in any case fully compatible with the phenomenological constraints.
This is true also for the other selected EOS,
with the possible exception of the DBHF one,
which appears too repulsive at higher density.
The analysis indicates that the EOS at low density must be relatively soft.

\begin{figure}
\includegraphics[width=0.96\hsize,clip=true]{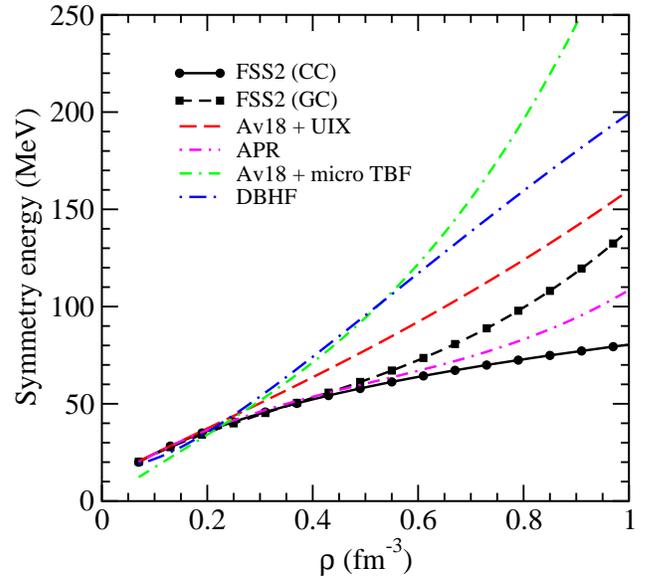}
\caption{(Color online)
The symmetry energy for several EOS. }
\label{fig:sym_en}
\end{figure}

Another gross property of the nuclear EOS,
which plays a decisive role in NS calculations,
is the symmetry energy as a function of density $S(\rho)$,
especially at the high density typical of the NS inner core.
It can be expressed in terms of the energy per particle between PNM and SNM,
\be
 \esym(\rho) = -{1\over4} {\partial(E/A) \over \partial x}(\rho,0.5)
 \approx {E\over A}(\rho,0) - {E\over A}(\rho,0.5) \:,
\label{e:sym}
\ee
and is reported in Fig.~\ref{fig:sym_en} for the considered set of EOS.
A large spread of values is present for densities above saturation.
In comparison with the other EOS,
the fss2 EOS appears in the region of lower values (``iso-soft" EOS),
but the gap and continuous choices show an appreciable discrepancy
at higher density.
As we will see,
the stiffness of NS matter shows a reduced spread of values,
since the beta-equilibrium condition produces a compensation effect due to
the interplay of the size of the symmetry energy and the stiffness of the EOS.

\begin{figure}
\includegraphics[width=0.96\hsize,clip=true]{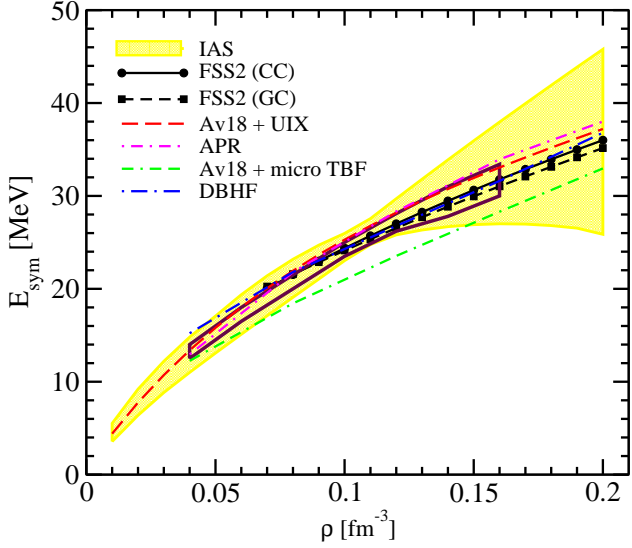}
\caption{(Color online)
The symmetry energy at low density.
The yellow band represents recent constraints \cite{Dani}
and the red line shows the region restricted by the neutron skin data,
whereas the different curves are
the results of the microscopic many-body methods.}
\label{fig:Dani}
\end{figure}

The symmetry energy up to saturation density has been constrained in
Ref.~\cite{Dani} by analyzing the data on isobaric analog states
as well as on the neutron skin in a set of nuclei.
In Fig.~\ref{fig:Dani} the larger (yellow) band indicates the constraint
coming from the analog states,
while the more restricted region bounded by the full (red) line is obtained
if also the neutron skin data are added.
The fss2 EOS is consistent throughout the constrained regions.
The other EOS look also consistent with the constraints,
with the possible exception of the BHF calculation with microscopic TBF.

\begin{figure}
\includegraphics[width=0.8\hsize,clip=true]{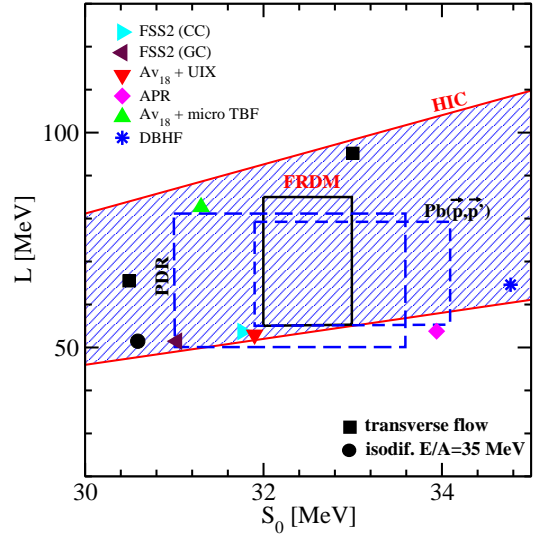}
\caption{(Color online)
$L=3\rho(\partial\esym/\partial\rho)|_{\rho=\rho_0}$
versus symmetry energy at saturation ${\esym}_0$,
predicted by several EOS.
See text for details of the experimental constraints.}
\label{fig:LS0}
\end{figure}

Another parameter which characterizes the symmetry energy
is its slope at saturation,
usually embodied in the quantity
$L \equiv 3\rho (\partial\esym/\partial\rho)|_{\rho=\rho_0}$.
In Fig.~\ref{fig:LS0} this parameter is displayed
versus the value of the symmetry energy at saturation
${\esym}_0\equiv\esym(\rho_0)$,
which has been widely discussed in Ref.~\cite{tsang2012}.
The different boxes indicate several constraint regions
obtained in different phenomenological analyses.
The dashed (blue) band represents the constraint coming from
experimental data on HIC,
obtained from the neutron and proton spectra from central collisions for
$^{124}$Sn+$^{124}$Sn and $^{112}$Sn+$^{112}$Sn reactions at 50 MeV/A \cite{fam}.
At the same incident energy, isospin diffusion was investigated.
We remind that isospin diffusion in HIC depends on the different
$N/Z$ asymmetries of the involved projectiles and targets,
hence it is used to probe the symmetry energy \cite{isodif1,isodif2,li}.
The full black circle shows the results from isospin diffusion observables
measured for collisions at a lower beam energy of 35 MeV per nucleon
\cite{isostar}.
Transverse collective flows of hydrogen and helium isotopes as well as
intermediate mass fragments with $Z<9$ have also been measured
at incident energy of 35 MeV/A in
$^{70}$Zn+$^{70}$Zn, $^{64}$Zn+$^{64}$Zn, $^{64}$Ni+$^{64}$Ni
reactions and compared to transport calculations.
The analysis yielded values denoted by the full black squares \cite{isotope}.

The box labeled by FRDM (finite-range droplet model)
represents a refinement of the droplet model \cite{mol},
and includes microscopic ``shell" effects and the extra binding
associated with $N=Z$ nuclei.
The FRDM reproduces nuclear binding energies of known nuclei within $0.1\%$,
and allows determination of both
${\esym}_0=32.5 \pm 0.5$ MeV and $L=70 \pm 15$ MeV.

The other boxes in Fig.~\ref{fig:LS0} represent experimental data obtained
from measurements of the neutron skin thickness.
In light nuclei with $N \approx Z$,
the neutrons and protons have similar density distributions.
With increasing neutron number $N$,
the radius of the neutron density distribution becomes larger than that
of the protons,
reflecting the pressure of the symmetry energy.
The measurement of the neutron skin thickness is made on the
stable nucleus $^{208}$Pb,
which has a closed neutron shell with $N=126$ and a closed proton shell
with $Z=82$,
hence it is very asymmetric and the neutron skin is very thick.
The possibility of measurements of the neutron radius in $^{208}$Pb by the
experiment PREX at Jefferson Laboratory has been widely discussed \cite{horo}.
The experiment should extract the value of the neutron radius in $^{208}$Pb
from parity-violating electron scattering.
However, the experimental signature is very small,
and the extracted thickness has a large statistical uncertainty.
In the next few years,
a second experimental run for PREX could reduce this large uncertainty
\cite{prex}.

Recent experimental data obtained by Zenihiro et al.~\cite{zeni} on the
neutron skin thickness of $^{208}$Pb deduced a value of
$\delta R_{np} = 0.211^{+0.054}_{-0.063}\;\text{fm}$.
From the experiments constraints on the symmetry energy were derived,
and these are plotted in Fig.~\ref{fig:LS0} as the
short-dashed blue rectangular box labeled Pb($\vec p,\vec p'$).

Last, we mention the experimental data on the Pygmy Dipole Resonance (PDR)
in very neutron-rich nuclei such as $^{68}$Ni and $^{132}$Sn,
which peaks at excitation energies well below the Giant Dipole Resonance,
and exhausts about $5\%$ of the energy-weighted sum rule \cite{pdr}.
In many models it has been found that this percentage is linearly dependent on
the slope $L$ of the symmetry energy.
Values of $L=64.8 \pm 15.7$ MeV and ${\esym}_0=32.2 \pm 1.3$ MeV
were extracted in Ref.~\cite{carb},
using various models which connect $L$ with the neutron skin thickness.
Those constraints are shown in Fig.~\ref{fig:LS0}
as a long-dashed rectangle with the label PDR.

It is not clear to what extent all these constraints are compatible
among each other,
but it looks that most of the EOS provide values consistent with
the general trend,
including the fss2 EOS.

\begin{figure}
\includegraphics[width=0.97\hsize,clip=true]{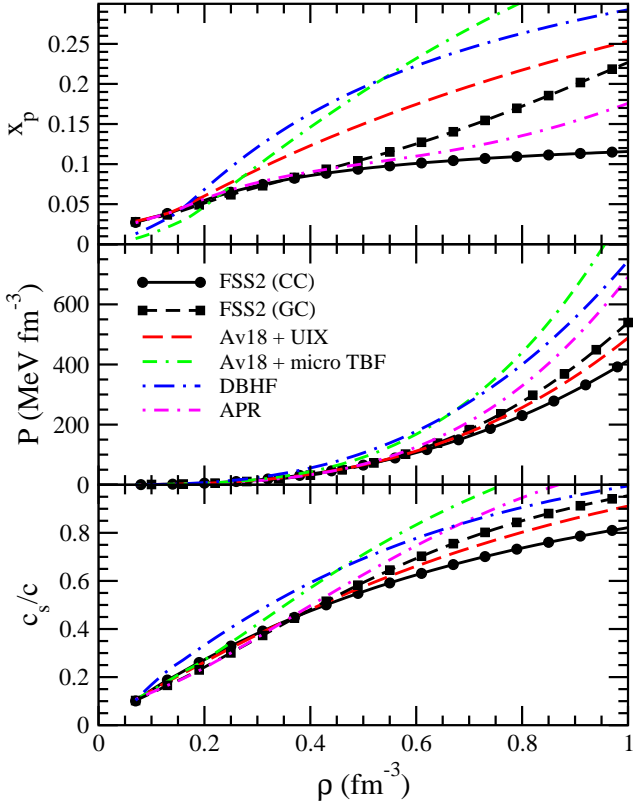}
\vskip-2mm
\caption{(Color online)
Proton fraction $x_p$ (upper panel),
pressure $P$ (middle panel),
and speed of sound in units of $c$ (lower panel)
of beta-stable nuclear matter for several EOS
as a function of baryon density~$\rho$.}
\label{fig:comp}
\end{figure}

\section{Neutron star structure}
\label{s:ns}

In order to study the structure of neutron stars,
we have to calculate the composition and the EOS of cold, neutrino-free,
catalyzed matter.
We require that the neutron star contains charge-neutral matter
consisting of neutrons, protons, and leptons ($e^-$, $\mu^-$)
in beta equilibrium,
and compute the EOS in the following standard way \cite{bbb,shapiro}:
The Brueckner calculation yields the energy density of
baryon/lepton matter as a function of the different partial densities,
\bea
 \eps(\rho_n,\rho_p,\rho_e,\rho_\mu) &=&
 (\rho_n m_n +\rho_p m_p)
 + (\rho_n+\rho_p) \frac{E}{A}(\rho_n,\rho_p)
\nonumber\\ &&
 +\, \rho_\mu m_\mu + {1\over 2m_\mu}{(3\pi^2\rho_\mu)^{5/3} \over 5\pi^2}
\nonumber\\ &&
 +\, { (3\pi^2\rho_e)^{4/3} \over 4\pi^2} \:,
\label{e:epsnn}
\eea
where we have used ultrarelativistic and nonrelativistic approximations
for the energy densities of electrons and muons, respectively.
In this study, we adopted the parabolic approximation for $E/A$,
Eq.~(\ref{e:parab}).
Knowing the energy density Eq.~(\ref{e:epsnn}),
the various chemical potentials (of the species $i=n,p,e,\mu$)
can be computed straightforwardly,
\be
 \mu_i = {\partial \eps \over \partial \rho_i} \:,
\ee
and the equations for beta equilibrium,
\be
 \mu_i = b_i \mu_n - q_i \mu_e \:,
\ee
($b_i$ and $q_i$ denoting baryon number and charge of species $i$)
and charge neutrality,
\be
 \sum_i \rho_i q_i = 0 \:,
\ee
allow one to determine the equilibrium composition $\{\rho_i\}$
at given baryon density $\rho$ and finally the EOS,
\be
 P(\rho) = \rho^2 {d\over d\rho}
 {\eps(\{\rho_i(\rho)\})\over \rho}
 = \rho {d\eps \over d\rho} - \eps
 = \rho \mu_n - \eps \:.
\ee

In Fig.~\ref{fig:comp} we report the proton fraction $x_p$,
the pressure $P$,
and the sound velocity,
\be
 c_s = \sqrt{\frac{\partial P}{\partial\eps}} \:,
\ee
as a function of total baryon density in NS matter.
The sound velocity can be used as a further test of a given EOS,
since it should not exceed the speed of light $c$.
One can see that the fss2 EOS is becoming superluminal only at very high density,
which, as we will see,
is actually not reached in the corresponding NS structure.

Once the EOS of the nuclear matter which is present throughout the NS is known,
one can use the Tolman-Oppenheimer-Volkoff \cite{shapiro,TOV1,TOV2} equations
for spherically symmetric NS:
\bea
 \frac{dp}{dr} &=& -\frac{Gm\eps}{r^2}
 \frac{ (1+p/\eps) (1+4\pi r^3p/m) }{1-2Gm/r} \:,
\\
 \frac{dm}{dr} &=& 4 \pi r^2\eps \:,
\eea
where $G$ is the gravitational constant and
$m(r)$ is the enclosed mass within a radius $r$.
Given a starting density $\eps_c$,
one integrates these equations until the surface $r=R$,
and the gravitational mass is obtained by $M_G=m(R)$.
The EOS needed to solve the TOV equations is taken from the
nuclear matter calculations as discussed above for the liquid-core region
and matched with the crust EOS,
which has been taken from Refs.~\cite{nv,fey,bps}.
This matching occurs at about two thirds of the saturation density,
where the EOS analytical fits of Eq.~(\ref{eq:Polyfit}) are still accurate.

\begin{figure}
\includegraphics[width=0.99\hsize]{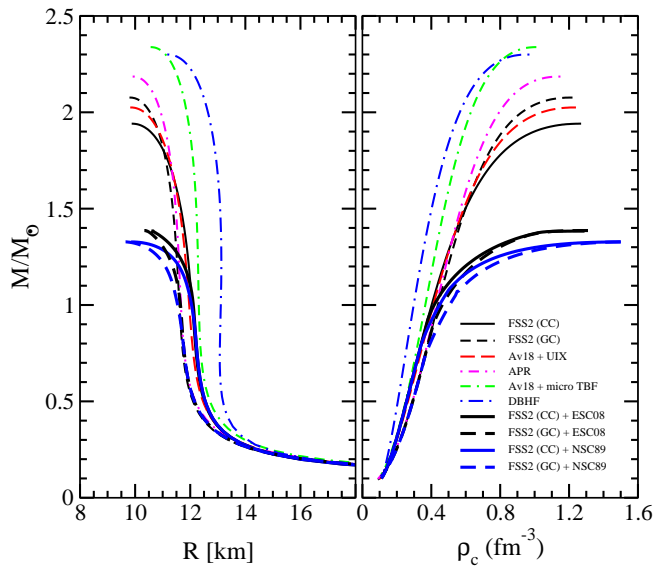}
\vskip -3mm
\caption{(Color online)
Neutron star mass as a function of radius (left panel)
or central baryon density (right panel)
for several EOS.
Thin lines indicate results obtained with purely nucleonic EOS,
whereas thick lines show results for several EOS including hyperons.
See text for details.}
\label{fig:MR}
\end{figure}

As is well known,
the mass of the NS has a maximum value as a function of radius
(or central density),
above which the star is unstable against collapse to a black hole.
The value of the maximum mass depends on the nuclear EOS,
so that the observation of a mass higher than the
maximum mass allowed by a given EOS simply rules out that EOS.
The fss2 EOS gives slightly different maximum masses
for the gap choice and continuous cases,
in line with their different stiffness at high density.
This gives a range of uncertainty for the maximum mass,
which encompasses the largest mass observed up to now,
which is $(2.01\pm0.04) M_\odot$ \cite{Anton}.
This is illustrated in Fig.~\ref{fig:MR},
where the relation between mass and radius (left panel)
or central density (right panel)
is displayed in comparison with the other considered EOS (thin lines).
The observed trend of the central density for all the EOS
is consistent with the corresponding $P(\rho)$ relation
displayed in the central panel of Fig.~\ref{fig:comp}.
As expected, with increasing incompressibility,
the NS central density decreases for a given mass.

In the end we illustrate the so-called ``hyperon puzzle''
with the fss2 model.
Fig.~\ref{fig:MR} (thick lines) shows the effect of allowing the appearance
of hyperons in beta-stable matter within our BHF approach \cite{H1,H2}.
Two different nucleon-hyperon (NY) interactions,
the Nijmegen NSC89 model \cite{nsc89} and the recent ESC08 model \cite{esc08}
are considered,
and combined with the fss2 NN potential in the approximate way explained
in more detail in Ref.~\cite{H1},
namely the purely nucleonic BHF energy density obtained with the fss2
is combined with the hyperonic contribution to the energy density
evaluated with either the NSC89 or the ESC08 interaction,
but together with the Argonne $V_{18}$ potential plus nucleonic TBF.
In this way the intermediate states in the NY Bethe-Goldstone equation
are treated approximately,
but the overall error of the global results is expected to be small \cite{H1}.

Fig.~\ref{fig:MR} demonstrates that under these assumptions the NS maximum
mass is practically insensitive to the choice of the NN interaction,
but determined by the NY interaction.
This is due to a well-known compensation mechanism
that can be clearly seen in the left panel of the figure:
The slightly stiffer fss2 gap-choice model
(dashed black and blue curves)
causes an earlier onset of hyperons and a stronger softening
than the fss2 continuous-choice model
(solid curves).
In any situation the maximum mass
is much smaller than current observational values.

\section{Summary and Discussion}
\label{s:end}

We have derived the nuclear matter EOS within the BBG
approach up to the THL level of approximation,
starting from an NN interaction
based on quark-quark and quark-meson interactions. 
An intrinsic uncertainty in the approach is related
to the choice of the auxiliary s.p.~potential.
Within this uncertainty the saturation point is well reproduced
without any additional parameters with respect to the interaction,
that is able also to reproduce the binding of three and four nucleon systems.
At higher energy the interaction should be improved in some channels,
in particular the $^{3}P_2$-$^{3}F_2$,
which is relevant for the high-density part of the EOS,
and therefore for NS.

The symmetry energy as a function of density up to saturation,
its value and slope at saturation,
and the incompressibility of symmetric matter at saturation
compare favorably well with the phenomenological constraints.
Above saturation the EOS is compatible with the flow data in
HIC at intermediate energy, up to about four times saturation density.

As already discussed,
a warning about these calculations is the observation that two and
three hole-line diagram contributions become comparable at higher density,
which puts some doubt on the convergence of the BBG expansion.
However, this is mainly due to the behavior of the two hole-line contribution,
which saturates or even decreases at higher density,
due to the compensation between positive and negative contributions.
This indicates
that the degree of convergence cannot be estimated in a straightforward way.
Within the present many-body treatment this is
probably the main source of uncertainty in the results.

However, let us notice that up to few times saturation
density the EOS calculated with the continuous and gap choice agree very well.
In particular for symmetric matter the
agreement extends up to the maximal density used in NS calculations.
This fact can be considered a good
indication for the convergence of the expansion,
because this agreement would be exactly true if convergence is indeed reached.
For similar reasons the mass-radius
relationship in NS are quite close for the two choices,
which gives support to the validity of these results.

The EOS can be considered relatively soft,
but despite that the NS maximum mass is compatible
with the current observed NS maximum mass of about 2 solar masses \cite{Anton}.
Phenomenology seems then to validate this microscopically derived EOS,
at least up to few times saturation density.

However, there are some theoretical caveats to be considered.
It can be expected that quark matter appears in the center of massive NS.
To describe these ``hybrid" NS one needs to know the quark matter EOS.
It turns out that many models for the deconfined quark matter produce
a too soft EOS to support a NS of mass compatible with observations
\cite{bag1,bag2,NJL,CDM,bag3,FCM,DS}.
The quark-quark interaction in the deconfined phase must be then
repulsive enough to stiffen the EOS, and indeed,
with a suitable quark-quark interaction,
mixed quark-nucleon matter can have an EOS compatible with two solar masses
or more \cite{Alford2013,Zuo}.

An additional problem arises if strange matter is introduced in the NS matter.
It turns out that BHF calculations using realistic hyperon-nucleon interactions
known in the literature produce a too soft NS matter EOS and the maximum mass
is reduced to well below the observational limit \cite{H1,H2}.
Although the hyperon-hyperon interactions
and in particular hyperonic TBF are poorly known,
these results pose a ``hyperon puzzle".
EOS based on relativistic mean field models
can solve the problem \cite{Haensel,Voskre,Oert}
with a proper choice of parameters.
Also modifications of the hyperon-nucleon interaction,
including three-body forces,
could provide a remedy for the too soft EOS \cite{Lona,Yama}.
All these methods introduce quark-quark or nucleon-hyperon and
hyperon-hyperon interactions that stiffen enough the EOS at high density.

In this respect it would be of particular relevance to perform BBG calculations
up to THL level with the quark-model baryon-baryon
interaction fss2 extended to the strange sector \cite{fss2YN}.
This difficult problem must be left to a future long-term project.

\acknowledgments

The authors wish to acknowledge the ``NewCompStar'' COST Action MP1304.
One of them (K.F.) would like to express his gratitude to Prof.~T.~Rijken
for providing him with the Nijmegen PSA data and the program
for the $\chi^2$ analysis.

\appendix
\section{Deuteron wave function}
\label{s:deut}

In this appendix,
we basically follow the notation of Ref.~\cite{fss2NN}.
First we solve the Lippmann-Schwinger RGM equation for the deuteron
\be
 (\gamma^2+k^2) f_\ell(k) = -{M_N}\frac{4\pi}{(2\pi)^3}
 \sum_{\ell'} \int^\infty_0 dq\,q^2 \,
 V_{\ell\ell'}(k,q) f_{\ell'}(q) \:.
\ee
Here we use the nonrelativistic expression
$\eps_d = \gamma^2/M_N$ as the deuteron binding energy.
The relativistic correction is of order $\eps_d^2/M_N$,
which corresponds to a few keV difference \cite{CD-Bonn}.
The total wave function is
\be
 \Psi_d^{1M}(\kv) =
 \left[ f_0(k){\cal Y}^{1M}_{01}(\hat\kv)
      + f_2(k){\cal Y}^{1M}_{21}(\hat\kv) \right] \zeta^0_0 \:,
\ee
where ${\cal Y}^{JM}_{LS}(\hat\kv)$ are spin-spherical harmonics
and the isospin function is denoted by $\zeta_T^{M_T}$.
In coordinate space, we have

\setlength{\tabcolsep}{0pt}
\begin{table}[ht]
\caption{
The range parameters $\gamma_j$ and
coefficients $C_j$ and $D_j$ for the parametrized deuteron wave function.
The values in parentheses are calculated from the boundary conditions.}
\begin{ruledtabular}
\begin{tabular}{dddd}
 j &   \gamma_j & C_j                   & D_j \\
\hline
 1 & 0.2314 &  0.8803           &  0.2218\times10^{-1} \\
 2 & 1.1314 & -0.2235           & -0.4637            \\
 3 & 2.0314 & -0.2501           &  0.1776            \\
 4 & 2.9314 & -0.1658\times10^2 & -0.1294\times10^2  \\
 5 & 3.8314 &  0.1694\times10^3 &  0.1192\times10^3  \\
 6 & 4.7314 & -0.9160\times10^3 & -0.5884\times10^3  \\
 7 & 5.6314 &  0.2592\times10^4 &  0.1611\times10^4  \\
 8 & 6.5314 & -0.4095\times10^4 & -0.2503\times10^4  \\
 9 & 7.4314 &  0.3678\times10^4 &(0.2214\times10^4) \\
10 & 8.3314 & -0.1766\times10^4 &(-0.1043\times10^4) \\
11 & 9.2314 & (0.3530\times10^3)&(0.2040\times10^3) \\
\end{tabular}
\end{ruledtabular}
\label{deut-para}
\end{table}

\be
 \Psi_d^{1M}(\rv) =
 \left[ (u_0(r)/r){\cal Y}^{1M}_{01}(\hat\rv)
       +(u_2(r)/r){\cal Y}^{1M}_{21}(\hat\rv) \right] \zeta^0_0 \:.
\ee
They are related by the Fourier transform
\be
 u_{\ell}(r) = \sqrt{2/\pi}\; i^\ell
 \int^\infty_0 dk\,k^2 r j_\ell(kr) f_\ell(k) \:.
\ee
The normalization is
\be
 \sum_{\ell=0,2} \int^\infty_0 dr\; u_\ell^2(r) =
 \sum_{\ell=0,2} \int^\infty_0 dk\,k^2 f_\ell^2(k) = 1 \:.
\ee

The deuteron wave functions are parametrized in the following way
as in Refs.~\cite{fss2NN} and \cite{CD-Bonn}
\bea
 f_{\ell\alpha}(k) &=& \sum^n_{j=1}
 \left\{\begin{array}{l} C_j \\ D_j \end{array}\right\}
 \sqrt{\frac{2}{\pi}} \frac{1}{k^2+\gamma_j^2}
 \hspace{3mm}\hbox{for }
\left\{\begin{array}{l} \ell=0 \\ \ell=2 \end{array}\right. \:,
\\
 u_{\ell\alpha}(r) &=& \sum^n_{j=1}
 \left\{\begin{array}{l}
 C_j e^{-\gamma_j r} \\
 D_j e^{-\gamma_j r}
 \left( 1+\frac{3}{\gamma_jr}+\frac{3}{(\gamma_jr)^2} \right)
 \end{array}\right.
 \hspace{0mm}\hbox{for }
 \left\{\begin{array}{l} \ell=0 \\ \ell=2 \end{array}\right. \:.
\nonumber\\&&
\eea
The range parameters are chosen as
$\gamma_j = \gamma + (j-1)\gamma_0$
with $\gamma_0 = 0.9\;\text{fm}^{-1}$ and $n=11$.

For $r\rightarrow\infty$,
the deuteron wavefunctions have the form
$u_0(r)$ $\rightarrow A_S e^{-\gamma_1 r}$ and
$u_2(r) \rightarrow A_D e^{-\gamma_1 r}
\left[ 1 + 3/(\gamma_1 r) + 3/(\gamma_1r)^2 \right]$,
where $A_S=C_1$ and $A_D=D_1$.
The asymptotic $D$-state to $S$-state ratio is given by $\eta = A_S/A_D$.
The boundary conditions $u_0(r) \rightarrow r$ and $u_2(r) \rightarrow r^3$
as $r \rightarrow 0$
lead to one constraint for the $C_j$ and three constraints for the $D_j$
\cite{PLB.101.139.(1981)}.
These constraint for the last $C_j$ and the last three $D_j$
are explicitly written in Eqs.~(C.7) and (C.8) of Ref.~\cite{CD-Bonn}.
The values of $\gamma_j$, $C_j$, and $D_j$ are listed in Tab.~\ref{deut-para}.
The accuracy of the parametrization is characterized by
\bea
 \left\{ \int^\infty_0 dk\,k^2
 \left[ f_0(k)-f_{0\alpha}(k) \right]^2 \right\}^{1/2} &=& 1.9\times10^{-4} \:,
\\
 \left\{ \int^\infty_0 dk\,k^2
 \left[ f_2(k)-f_{2\alpha}(k) \right]^2 \right\}^{1/2} &=& 2.3\times 10^{-4} \:.
\eea
The quadrupole moment $Q_d$,
the root mean square radius $R_d$,
and the $D$-state probability $P_D$
are calculated using Eqs.~(C.16)-(C.18) of Ref.~\cite{CD-Bonn}, respectively.


\end{document}